\documentclass[aps,prapplied,twocolumn,superscriptaddress,longbibliography]{revtex4-2}

\usepackage{amsmath,amssymb,bm,siunitx,physics}
\usepackage{graphicx}
\usepackage{booktabs}
\usepackage[hidelinks]{hyperref}
\usepackage[capitalise]{cleveref}
\usepackage{microtype}
\usepackage{booktabs}
\begin{document}

\title{Internal Charge Amplification in Germanium at 77\,K and 4\,K:\\ From Single-Free-Flight Bounds to a Physics-Informed Ionization Model}

\author{Dongming Mei}
\affiliation{Department of Physics, University of South Dakota, Vermillion, SD 57069, USA}
\email{Dongming.Mei@usd.edu}
\author{Kunming Dong}
\affiliation{Department of Physics, University of South Dakota, Vermillion, SD 57069, USA}
\author{Narayan Budhathoki}
\affiliation{Department of Physics, University of South Dakota, Vermillion, SD 57069, USA}
\author{Shasika Panamaldeniya}
\affiliation{Department of Physics, University of South Dakota, Vermillion, SD 57069, USA}
\author{Francisco Ponce}
\affiliation{National Security Directorate, Pacific Northwest National Laboratory, Richland, Washington, 99354, USA  }
\date{\today}

\begin{abstract}
Internal charge amplification (ICA) in cryogenic high-purity germanium (HPGe) can lower detection thresholds by providing gain inside the detector crystal, but reliable operation requires a predictive estimate of the avalanche-onset \emph{critical electric field} \(E_{\mathrm{crit}}\). We present a compact framework for \(E_{\mathrm{crit}}\) at 77~K and 4~K (typical HPGe operating temperatures) that bridges (i) a mobility-based single-free-flight (SFF) upper bound with (ii) a physics-informed impact-ionization model incorporating energy-dependent scattering, nonparabolic (Kane) dispersion, intervalley transfer, and the high-energy ``lucky-drift'' tail. This unified treatment yields closed-form, design-useful relations, including \(E_{\mathrm{crit}}^{(\mathrm{PI})}=B(T)/\ln[A(T)d]\), and a practical calibration workflow that maps measured low-field mobility \(\mu(T)\) and gain curves \(M(V)\) (Chynoweth analysis) to device-level bias targets with propagated uncertainty bands. Example electron and hole estimates indicate that realistic transport typically lowers \(E_{\mathrm{crit}}\) relative to SFF and increases the predicted change in \(E_{\mathrm{crit}}\) between 77~K and 4~K. The resulting portable formulas connect materials/transport inputs to geometry, excess noise, and field shaping, providing design-ready guidance for stable, unipolar-favored ICA with controlled quenching in Ge and other cryogenic semiconductors.
\end{abstract}

\maketitle

\section{Introduction}
Internal charge amplification (ICA), often called avalanche multiplication, occurs when charge carriers accelerated by an electric field gain sufficient energy between inelastic scattering events to \emph{impact ionize}, generating secondary electron--hole pairs and producing gain in the charge-collection channel~\cite{SzeNg,Chynoweth1958,OkutoCrowell1974}. In high-purity germanium (HPGe) operated at cryogenic temperatures (77~K and 4~K), long carrier mean free paths, a small band gap, and mature control of defects and contacts make Ge a compelling platform for low-noise ICA. For practical detector design, however, ICA must be predictable and stable: a key quantity is the \emph{critical electric field} \(E_{\mathrm{crit}}\), defined here as the onset field for sustained internal amplification under device-relevant conditions. Reliable estimates of \(E_{\mathrm{crit}}\) are essential for selecting bias targets, shaping fields, and avoiding premature breakdown.

\paragraph*{Motivation from low-mass dark matter (LDM).}
A major driver for ICA in Ge is sensitivity to \(\mathcal{O}(1\text{--}1000)\,\mathrm{MeV}\) dark matter, where both electron-recoil (ER) and nuclear-recoil (NR) signatures lie at the eV--sub-keV scale~\cite{Essig2012,Essig2016,Billard2014}. Sub-GeV DM--electron scattering produces only a few electron--hole pairs in semiconductors, while low-mass DM--nucleus scattering yields \(\lesssim\!\mathrm{keV}\) nuclear recoils. In this regime, competitive cross-section reach demands \emph{ultra-low thresholds} with high-efficiency triggering and robust background rejection. Recent results highlight the trend: cryogenic semiconductor experiments (e.g., SuperCDMS~\cite{SuperCDMSSNOLAB2017}) aim for eV-scale phonon-mediated thresholds, and CRESST-III has demonstrated \(\sim\!10\,\mathrm{eV}\) thresholds in dedicated targets~\cite{Angloher2023}. ICA can complement phonon-based amplification by providing \emph{in-sensor} charge gain, improving front-end signal-to-noise and enabling triggering on single- or few-pair events while preserving charge-based energy reconstruction in Ge. Related concepts also exploit internal amplification for ionization from impurity states to access MeV-scale dark-matter signatures~\cite{Mei2018EPJC,Wei2022PPPCGe}.

\paragraph*{Motivation from coherent elastic neutrino--nucleus scattering (CE\(\nu\)NS).}
CE\(\nu\)NS, proposed by Freedman~\cite{Freedman1974} and first observed by COHERENT~\cite{COHERENT2017}, produces keV and sub-keV nuclear recoils for \(\mathcal{O}(\mathrm{MeV})\) neutrinos from spallation sources and reactors. The physics program includes precision tests of the weak interaction, searches for non-standard interactions, neutrino magnetic moments and sterile neutrinos, and applications such as reactor monitoring. For Ge-based CE\(\nu\)NS detectors, sensitivity is limited primarily by \emph{energy threshold} and \emph{background control}. ICA addresses the former by amplifying the primary ionization prior to readout, improving trigger efficiency at fixed exposure and relaxing front-end noise constraints. Combined with cryogenic operation, Ge ICA offers a practical route toward scalable CE\(\nu\)NS measurements with thresholds in the few--tens of eV regime~\cite{Billard2014,COHERENT2017}.

\paragraph*{Technical challenges.}
Achieving stable, low-noise ICA in Ge requires simultaneously managing materials, device electrostatics, and readout. Device-level multiplication demands \emph{high fields that are also uniform} and free of localized hot spots that trigger microplasma breakdown; this depends sensitively on curvature, sidewall finish, guard structures, and contact technologies (e.g., amorphous-Ge versus Li-diffused contacts) that also affect surface leakage and dead layers~\cite{SzeNg}. At cryogenic temperatures, free-flight statistics and the high-energy carrier tail are shaped by ionized- and neutral-impurity scattering, intervalley phonons, and intervalley transfer, which feed directly into impact-ionization rates and gain stability~\cite{Jacoboni1983,Conwell1967,Chynoweth1958,OkutoCrowell1974}. On the readout side, thresholds are constrained not only by front-end noise but also by the \emph{excess noise} intrinsic to stochastic multiplication, so geometry and field shaping must balance achievable gain against noise, bandwidth, and cryogenic heat load~\cite{SzeNg,OkutoCrowell1974}. Finally, calibration at the \(\mathcal{O}(\mathrm{eV})\)–sub-keV scale is required to validate energy response and stability over temperature cycles, and to quantify systematic shifts in \(\alpha(E,T)\) driven by nonparabolicity and energy-dependent scattering~\cite{Jacoboni1983,Conwell1967}. A practical first step is therefore to predict the \textbf{critical field} \(E_{\mathrm{crit}}\) and to engineer geometries and contact layouts that realize it reproducibly while remaining below breakdown hot spots and within readout noise budgets.

\paragraph*{Modeling challenge and our approach.}
Existing approaches for avalanche onset in Ge and related semiconductors typically fall into two categories. At one extreme, single--free-flight (SFF) estimates equate field-induced energy gain over a momentum-relaxation time \(\tau\) to an effective ionization threshold \(\varepsilon_i\). These yield transparent mobility-based \emph{upper bounds} on \(E_{\mathrm{crit}}\) but do not capture the physics controlling the \emph{high-energy tail}: energy-dependent scattering (acoustic/optical/intervalley phonons, ionized/neutral impurities), nonparabolic band dispersion, and intervalley transfer~\cite{Kane1957,Jacoboni1983,Conwell1967}. At the other extreme, full-band Monte Carlo simulations and purely empirical Chynoweth fits can reproduce measured ionization coefficients, but they are cumbersome for early-stage geometry optimization and often do not provide compact, design-ready relations for \(E_{\mathrm{crit}}\) in cryogenic HPGe devices.

Here we bridge these viewpoints and provide a compact, physics-informed (PI) design framework that links microscopic transport to device-level breakdown criteria. First, we formalize the SFF scaling that ties \(E_{\mathrm{crit}}\) to the measured mobility \(\mu(T)\) and an effective \(\varepsilon_i\), clarifying its role as an \emph{upper bound} in cryogenic HPGe. Second, starting from Kane-type nonparabolic dispersion and energy-dependent inelastic scattering, we derive an analytic form for the ionization coefficient,
\(\alpha(E,T) \simeq A(T)\exp[-B(T)/E]\), where \(B(T)\) emerges from an explicit energy-relaxation integral over the high-energy tail and \(A(T)\) parameterizes post-threshold ionization. Third, applying a device-level onset criterion \(\int_0^d \alpha(E,T)\,dx \simeq 1\) for multiplication width \(d\) yields a closed-form critical field,
\begin{equation}
E_{\mathrm{crit}}^{(\mathrm{PI})} = \frac{B(T)}{\ln[A(T)d]}.
\end{equation}
This result provides a practical design rule that connects \(E_{\mathrm{crit}}\) to geometry and to parameters \((A,B)\) that can be calibrated from short, uniform-field test diodes (e.g., via Chynoweth analysis), while maintaining a clear link to mobility-based bounds and transport physics relevant at 77~K and 4~K.

Compared with prior treatments that rely either on empirical Chynoweth fits or full-band Monte Carlo, this work makes three specific contributions. (i) We derive an explicit bridge between the mobility-based SFF bound and a PI impact-ionization model by expressing the Chynoweth parameter \(B(T)\) as an energy-relaxation integral controlled by band structure and inelastic scattering. (ii) We recast the transport model into a portable, device-level rule \(E_{\mathrm{crit}}^{(\mathrm{PI})}=B(T)/\ln[A(T)d]\) that connects microscopic transport to the multiplication width and can be calibrated from short diodes. (iii) We embed the framework in a Ge ICA case study using a realistic impurity gradient and a TCAD-computed field map, illustrating how the formulas guide geometry choices, bias window, and expected gain/noise behavior in HPGe devices.

The remainder of this paper is organized as follows. Section~\ref{sec:sff} introduces the SFF picture. Section~\ref{sec:physics-informed} develops the PI transport model and its mapping to the Chynoweth parameters \((A,B)\). Section~\ref{sec:device_level} formulates the device-level onset criterion, and Sec.~\ref{sec:device_design} applies the framework to an ICA-capable Ge detector geometry. In Sec.~\ref{sec:calibration} we outline a calibration and uncertainty-propagation workflow for extracting \((A(T),B(T))\) from short test diodes at 77~K and 4~K. Section~\ref{sec:discussion} provides a broader discussion and outlook, and Sec.~\ref{sec:conclusion} summarizes the resulting design rules and implications for next-generation cryogenic Ge detectors for low-threshold rare-event searches.

\section{SFF estimate}
\label{sec:sff}

The SFF picture provides a compact \emph{upper bound} for the onset of internal charge amplification by equating the field-induced kinetic-energy gain during one momentum-relaxation time to an effective impact-ionization threshold~\cite{SzeNg,Chynoweth1958,OkutoCrowell1974}. For a carrier undergoing a free flight of duration \(\tau\) in a uniform field \(E\), the acceleration is \(a=qE/m^\ast\), so the velocity gain is \(\Delta v = (qE/m^\ast)\tau\) and the corresponding kinetic-energy gain is
\begin{equation}
  \Delta\varepsilon
  \;=\;
  \frac{1}{2} m^\ast (\Delta v)^2
  \;=\;
  \frac{q^2 E^2 \tau^2}{2 m^\ast},
\end{equation}
where \(q\) is the carrier charge and \(m^\ast\) is the conductivity effective mass. Using the Drude relation \(\mu = q\tau/m^\ast\) to eliminate \(\tau\), and imposing \(\Delta\varepsilon=\varepsilon_i\) (an effective threshold for initiating impact ionization), yields the familiar SFF bound
\begin{equation}
  \boxed{E_{\mathrm{crit}}^{\mathrm{(SFF)}}=\frac{\sqrt{2\,\varepsilon_i/m^\ast}}{\mu}},
  \label{eq:SFF}
\end{equation}
which makes explicit the robust scalings \(E_{\mathrm{crit}}\propto \mu^{-1}\) and \(E_{\mathrm{crit}}\propto \sqrt{\varepsilon_i}\).
For example, if \(\mu(4~\mathrm{K})\approx 10\,\mu(77~\mathrm{K})\), then
\(E_{\mathrm{crit}}^{\mathrm{(SFF)}}(4~\mathrm{K})\approx E_{\mathrm{crit}}^{\mathrm{(SFF)}}(77~\mathrm{K})/10\).

\subsection{Assumptions and scope}

The SFF construction is deliberately minimalist. It assumes (i) a single, representative momentum-relaxation time \(\tau\) (energy independent), (ii) parabolic dispersion with a temperature-independent \(m^\ast\), and (iii) negligible in-flight inelastic loss up to \(\varepsilon_i\). These assumptions maximize the probability of reaching \(\varepsilon_i\) in one flight; consequently, \cref{eq:SFF} should be interpreted as an \emph{upper bound} on the practical onset field. In reality, energy-dependent scattering, nonparabolicity, and intervalley transfer suppress the high-energy tail relative to the SFF idealization~\cite{Jacoboni1983,OkutoCrowell1974}.

\subsection{Choice of impact threshold \(\varepsilon_i\)}

The effective threshold \(\varepsilon_i\) is not unique, and different choices are useful for bracketing design space. A pragmatic range lies between the band gap \(E_g\) and the mean pair-creation energy \(W\). For Ge, \(E_g\simeq 0.73~\mathrm{eV}\) and \(W\simeq 2.96~\mathrm{eV}\) at cryogenic temperatures~\cite{SzeNg}. Taking \(\varepsilon_i\simeq W\) is conservative (higher fields), while \(\varepsilon_i \simeq \eta E_g\) with \(\eta\sim 1\text{--}4\) reflects microscopic thresholds inferred from transport fits and avalanche data~\cite{OkutoCrowell1974}. Given this ambiguity, reporting \(E_{\mathrm{crit}}^{\mathrm{(SFF)}}\) as a band versus \(\varepsilon_i\) is good practice.

\subsection{Effective masses and anisotropy}

For electrons in Ge, high-field transport proceeds primarily in the L valleys; a representative conductivity mass \(m_e^\ast\!\approx\!0.12\,m_0\) is often used for first estimates. Holes exhibit heavier and more anisotropic masses, for which \(m_h^\ast\!\approx\!0.28\,m_0\) provides a useful single-parameter summary~\cite{SzeNg,Jacoboni1983}. Because \(E_{\mathrm{crit}}^{\mathrm{(SFF)}}\propto (m^\ast)^{-1/2}\), uncertainties in \(m^\ast\) propagate directly into the bound.

\subsection{Equivalent mean-free-path form}

Using a characteristic mean free path \(\ell = v\tau\) (with \(v\) taken as a representative drift or thermal velocity), the SFF criterion can be written as
\begin{equation}
  q E_{\mathrm{crit}}^{\mathrm{(SFF)}}\,\ell \;\simeq\; \varepsilon_i,
  \qquad
  \ell \equiv v\,\tau
  \;=\;
  \frac{m^\ast v}{q}\,\mu,
\end{equation}
which highlights a simple design rule: over one mean free path, the field must supply the energy needed to reach the impact-ionization threshold \(\varepsilon_i\).
In regimes approaching velocity saturation, replacing \(v\) by \(v_{\mathrm{sat}}\) yields a more realistic \(\ell\) and further reinforces the ``upper-bound'' character of the SFF estimate~\cite{Jacoboni1983}.

\subsection{Temperature and purity leverage}
Lowering the lattice temperature suppresses phonon populations, lengthening the momentum-relaxation time $\tau$ and increasing the mobility $\mu$. In the SFF bound the onset field scales inversely with mobility,
$E_{\mathrm{crit}}^{(\mathrm{SFF})}\propto 1/\mu$,
so any increase in $\mu$ (for example, cooling from 77~K to 4~K) produces a first-order reduction in the predicted SFF onset field without requiring a detailed accounting of the individual scattering channels~\cite{SzeNg}. Improving material purity acts in the same direction by reducing ionized/neutral impurity and defect scattering, thereby increasing $\tau$ and $\mu$.

For weakly doped Ge, commonly used empirical fits for the \emph{low-field Hall mobility} in the phonon-dominated range give
$\mu_n(T)\approx 4.9\times10^{7}T^{-1.66}$ for electrons (quoted for 77--300~K) and
$\mu_p(T)\approx 1.05\times10^{9}T^{-2.33}$ for holes (quoted for 100--300~K), with $\mu$ in $\mathrm{cm^2/(V\,s)}$ \cite{IoffeGeElectric}.
Interpreted strictly as \emph{upper-bound extrapolations} of phonon-limited transport, these fits imply
$\mu_n(77~\mathrm{K})\sim 3.6\times10^{4}$ and
$\mu_p(77~\mathrm{K})\sim 4.2\times10^{4}\ \mathrm{cm^2/(V\,s)}$.
Below $\sim$10~K, however, such extrapolations become unreliable: phonon scattering rapidly weakens and momentum relaxation becomes increasingly controlled by residual ionized impurities, neutral impurities, and defects, making the mobility strongly sample dependent~\cite{Mei2016NeutralPtypeGeJINST,Mei2017NeutralNtypeGeJINST,Mei2024ResidualImpuritiesGeDM}.

Direct cryogenic drift measurements provide an experimentally anchored reference at even lower temperatures. In ultrapure Ge, SuperCDMS measured electron and hole drift velocities at 31--50~mK as a function of applied field in the low-field regime relevant to cryogenic phonon/ionization detectors \cite{Sundqvist2009DriftVelocities,Phipps2012OpticallyInduced}. It is convenient to summarize these data with an \emph{effective} mobility,
\begin{equation}
\mu_{\mathrm{eff}}(E)\equiv \frac{v_d(E)}{E},
\end{equation}
which serves as a scalar proxy for inherently anisotropic transport (notably, electrons can propagate obliquely relative to $\vec E$ in Ge). Using the drift-speed points overlaid in Fig.~7 of \cite{Kelsey2023G4CMP} (from \cite{Phipps2012OpticallyInduced}), representative values at 50~mK are
$\mu_{e,\mathrm{eff}}\approx (2.6\pm0.2)\times10^{6}$ and
$\mu_{h,\mathrm{eff}}\approx (1.8\pm0.2)\times10^{6}\ \mathrm{cm^2/(V\,s)}$ near $E\simeq 1~\mathrm{V/cm}$,
decreasing to
$\mu_{e,\mathrm{eff}}\approx (1.1\pm0.1)\times10^{6}$ and
$\mu_{h,\mathrm{eff}}\approx (0.7\pm0.1)\times10^{6}\ \mathrm{cm^2/(V\,s)}$ near $E\simeq 3~\mathrm{V/cm}$.
The decrease of $\mu_{\mathrm{eff}}(E)$ with field reflects the onset of hot-carrier energy-loss mechanisms (including Neganov--Trofimov--Luke phonon emission), which produce a sublinear increase of $v_d(E)$ at sub-kelvin temperatures \cite{Kelsey2023G4CMP}. In the same temperature regime, dopants and compensating impurities largely freeze out into localized neutral configurations (dipolar states~\cite{Mei2024ResidualImpuritiesGeDM}), so that neutral-impurity and defect scattering can dominate momentum relaxation and further enhance sample-to-sample variability \cite{Mei2017NeutralNtypeGeJINST,Mei2016NeutralPtypeGeJINST}. Accordingly, the SuperCDMS millikelvin values should be regarded as an \emph{upper-end benchmark} for very clean material at low fields rather than a universal expectation.

At the few-kelvin scale, early Hall-effect measurements already established that high-purity Ge exhibits a steep rise in low-field mobility when cooled from 77~K into the liquid-helium regime. Koenig \emph{et al.} measured the Hall mobility in high-purity $n$-type Ge and showed that it follows the acoustic-phonon–limited trend $\mu\propto T^{-3/2}$ over a broad range (their reference line is normalized to the measured value at $77~\mathrm{K}$), with the cleanest samples remaining close to the lattice-limited behavior down to $\sim 6~\mathrm{K}$ \cite{Koenig1962,Morin1957}. Cryogenic transport calculations based on particle Monte-Carlo methods, which include ionized-impurity scattering together with inelastic phonon processes, reproduce the Ohmic regime and support the same physical picture: as $T$ decreases, phonon scattering is suppressed and the mobility rises until residual impurities and defects become the dominant limitation \cite{AubryFortuna2010}. These results motivate a conservative detector-grade estimate $\mu(4~\mathrm{K})\sim 10^{5}~\mathrm{cm^{2}\,V^{-1}\,s^{-1}}$, corresponding to an order-of-magnitude enhancement relative to typical 77~K values in high-purity material \cite{Koenig1962,AubryFortuna2010,Ottaviani1973,Szmulowicz1983}.

Given this range of reported low-temperature behavior, we adopt a practical design-level scaling
$\mu(4~\mathrm{K})\approx 10\,\mu(77~\mathrm{K})$
for the analyses below, and treat the factor of 10 as an order-of-magnitude guide whose precise value depends on the crystal’s residual neutral-impurity/defect budget and is best established by dedicated low-temperature Hall or time-of-flight measurements on the specific detector-grade material \cite{Bradley2020CryogenicGeThesis}.

\subsection{Sensitivity and quick-look numerics}

For rapid scoping, it is useful to tabulate SFF fields for electrons and holes as functions of \(\mu\), \(\varepsilon_i\), and \(m^\ast\). Using \(\varepsilon_i=2.96~\mathrm{eV}\), \(m_e^\ast=0.12\,m_0\), \(m_h^\ast=0.28\,m_0\), \(\mu_e(77~\mathrm{K})=3.6\times10^{4}~\mathrm{cm^2/V\,s}\), \(\mu_h(77~\mathrm{K})=4.2\times10^{4}~\mathrm{cm^2/V\,s}\), and \(\mu(4~\mathrm{K})=10\,\mu(77~\mathrm{K})\), the resulting SFF \emph{upper-bound} critical fields are:

\begin{table}[h]
\centering
\caption{SFF upper-bound critical fields in Ge for representative parameters.}
\label{tab:sff_summary}
\begin{tabular}{lcccccc}
\toprule
Carrier & $T$ (K) & $\mu$ ($\mathrm{cm^2/V\,s}$) & $m^\ast/m_0$ & $\varepsilon_i$ (eV) & $E_{\rm crit}$ (V/cm) \\
\midrule
e$^-$ & 77 & $3.6\times 10^{4}$ & 0.12 & 2.96 & $8.2\times 10^{3}$ \\
h$^+$ & 77 & $4.2\times 10^{4}$ & 0.28 & 2.96 & $4.6\times 10^{3}$ \\
e$^-$ & 4  & $3.6\times 10^{5}$ & 0.12 & 2.96 & $8.2\times 10^{2}$ \\
h$^+$ & 4  & $4.2\times 10^{5}$ & 0.28 & 2.96 & $4.6\times 10^{2}$ \\
\bottomrule
\end{tabular}

\vspace{4pt}
\footnotesize
\emph{Notes:} (i) \(E_{\rm crit}\propto \mu^{-1}\sqrt{\varepsilon_i}\); adopting a lower microscopic threshold (e.g., \(\varepsilon_i\sim E_g\)) scales entries by \(\sqrt{\varepsilon_i/2.96~\mathrm{eV}}\).
(ii) These values are \emph{upper bounds}; energy-dependent scattering and nonparabolic transport typically shift practical onset to lower fields (Sec.~\ref{sec:physics-informed}).
\end{table}

\begin{figure}[h]
\centering
\includegraphics[width=0.92\linewidth]{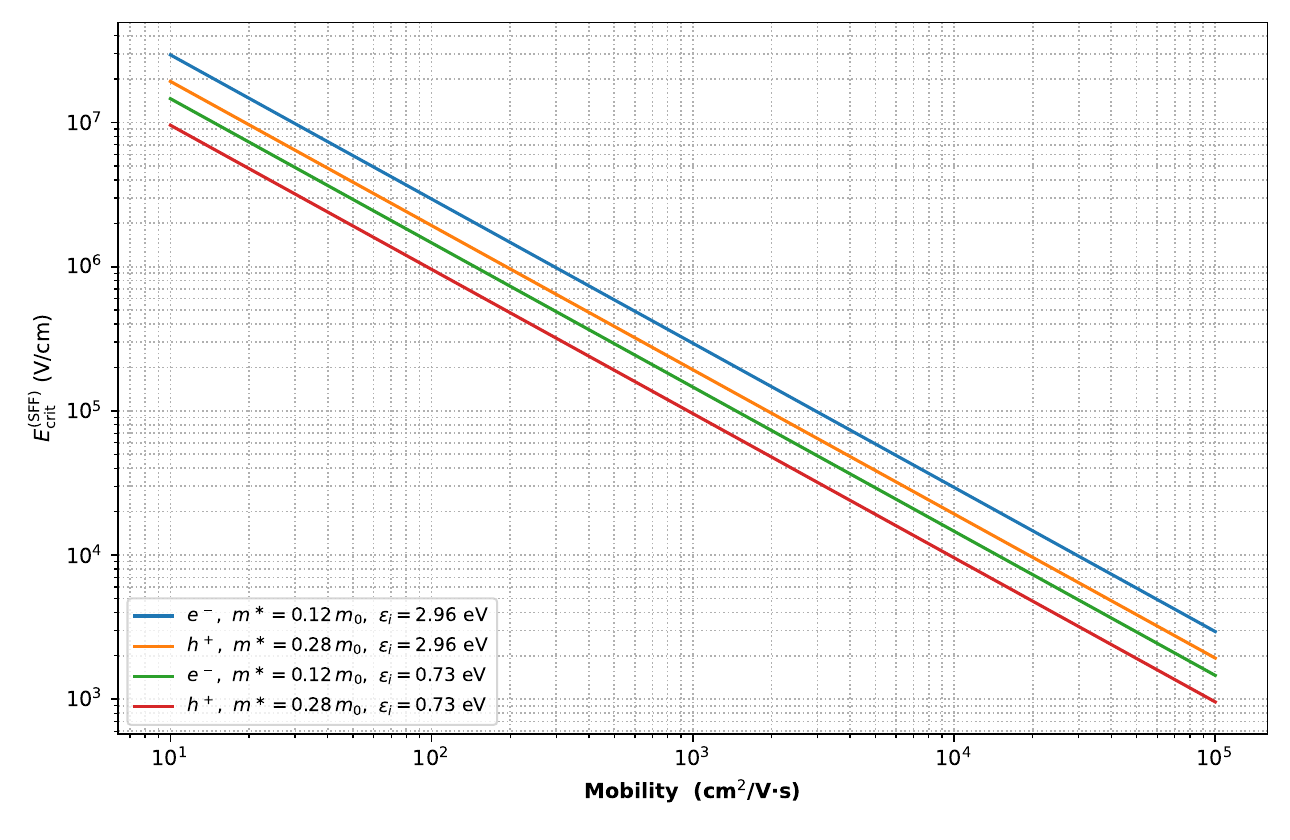}
\caption{Critical field \(E_{\mathrm{crit}}^{(\mathrm{SFF})}\) versus mobility \(\mu\) on log--log axes, comparing two thresholds \(\varepsilon_i=2.96~\mathrm{eV}\) (pair-creation energy) and \(\varepsilon_i=0.73~\mathrm{eV}\) (band gap), for electrons \((m^\ast=0.12\,m_0)\) and holes \((m^\ast=0.28\,m_0)\). The inverse scaling \(E_{\mathrm{crit}}\propto \mu^{-1}\) and \(\sqrt{\varepsilon_i}\) dependence are evident.}
\label{fig:Ecrit_vs_mu}
\end{figure}

\paragraph*{Visualization of scaling.}
We visualize these scalings in Fig.~\ref{fig:Ecrit_vs_mu}, which plots \(E_{\mathrm{crit}}^{(\mathrm{SFF})}\) versus mobility \(\mu\) on log--log axes for electrons \((m^\ast=0.12\,m_0)\) and holes \((m^\ast=0.28\,m_0)\). The curves show the expected \(E_{\mathrm{crit}}\!\propto\!\mu^{-1}\) behavior, with hole curves shifted upward by the larger effective mass. Comparing \(\varepsilon_i=2.96~\mathrm{eV}\) and \(0.73~\mathrm{eV}\) illustrates the \(\sqrt{\varepsilon_i}\) dependence as an approximately constant vertical offset across the full \(\mu\) range. These trends provide quick, order-of-magnitude targets for uniform-field design before applying the transport-informed treatment developed in Sec.~\ref{sec:physics-informed}.

\subsection{Limitations and connection to transport-informed models}
SFF bounds the field scale for “typical” carriers under simplified assumptions; tail physics and energy-dependent scattering shift practical onset.
Because real carriers undergo energy-dependent inelastic loss, nonparabolic dispersion, and intervalley transfer, the SFF criterion typically \emph{overestimates} the onset field. In transport-informed descriptions these effects enter through the ionization coefficient, often parameterized as \(\alpha(E)\simeq A\,\exp[-B/E]\), where \(B\) acts as an effective energy-relaxation scale for reaching the ionization threshold. In the parabolic/constant-\(\tau\) limit, the SFF bound is recovered as a special case, providing a useful calibration anchor for the physics-informed framework developed in Sec.~\ref{sec:physics-informed}~\cite{OkutoCrowell1974,Jacoboni1983}.

\section{PI transport and \texorpdfstring{\(\alpha(E)\)}{alpha(E)}}
\label{sec:physics-informed}

Building on the SFF bound in Sec.~\ref{sec:sff}, we now move from a
\emph{single-time, single-threshold} estimate to a transport-informed
description that explicitly tracks how carriers populate the
\emph{high-energy tail} under strong fields. In cryogenic Ge, avalanche onset
is controlled less by the mean energy gain per collision and more by the
probability that a carrier experiences an unusually long inelastic-free
trajectory (``lucky drift'') and reaches the ionization threshold before
losing energy to phonons or intervalley processes. This naturally leads to an
exponential ionization law \(\alpha(E,T)\simeq A(T)\exp[-B(T)/E]\), where
\(B(T)\) is an energy-relaxation integral that reduces to the SFF scale in the
parabolic, constant-\(\tau\) limit~\cite{Chynoweth1958,OkutoCrowell1974,Jacoboni1983}.

\subsection{Nonparabolic dispersion (Kane model)}

At high fields, carrier energies explore portions of the band where the
velocity--energy relation deviates from the parabolic form. A convenient
description is the Kane dispersion~\cite{Kane1957}:
\begin{align}
\varepsilon\bigl(1+\alpha_{\mathrm K}\varepsilon\bigr)&=\frac{\hbar^2 k^2}{2m_0^\ast},\\
v(\varepsilon)&=\frac{1}{\hbar}\frac{\partial\varepsilon}{\partial k}
=\sqrt{\frac{2\,\varepsilon\,(1+\alpha_{\mathrm K}\varepsilon)}{m_0^\ast\,(1+2\alpha_{\mathrm K}\varepsilon)}}\,,
\end{align}
with band-edge (conductivity) mass \(m_0^\ast\) and a nonparabolicity
parameter \(\alpha_{\mathrm K}\) (treated as a fit parameter tied to the
indirect gap and remote bands). Relative to the parabolic case, Kane
nonparabolicity lowers \(v(\varepsilon)\) at fixed \(\varepsilon\), which
\emph{increases} the energy-relaxation integral in Eq.~\eqref{eq:AB} and thus
drives a smaller PI-predicted onset field than the SFF bound for the same
\(\mu(T)\) and \(\varepsilon_i\)~\cite{Jacoboni1983,Conwell1967}.

\subsection{Energy-dependent scattering and intervalley transfer}

The inelastic scattering rate entering the energy-relaxation integral is
written as
\begin{equation}
\begin{aligned}
\nu_{\mathrm{inel}}(\varepsilon,T) &=
\nu_{\mathrm{ac}}(\varepsilon,T)
+\nu_{\mathrm{op}}^{\pm}(\varepsilon,T;\hbar\omega_{\mathrm{op}}) \\
&\quad
+\nu_{\mathrm{iv}}^{\pm}(\varepsilon,T;\hbar\omega_{\mathrm{iv}})
+\nu_{\mathrm{def}}(\varepsilon)\,,
\end{aligned}
\label{eq:energyrelaxing}
\end{equation}

collecting acoustic deformation-potential, optical phonons (emission/absorption),
intervalley phonons (e.g., L\(\leftrightarrow\)L transfers in Ge), and
defect/impurity-assisted inelastic channels~\cite{Jacoboni1983,Conwell1967}.
In Ge, \(\nu_{\mathrm{inel}}\) is dominated by deformation-potential acoustic
phonons at low energies, while optical and intervalley phonons become
important once \(\varepsilon\) exceeds their thresholds. At 77~K, both phonon
emission and absorption contribute, leading to efficient energy relaxation and
a comparatively larger \(\nu_{\mathrm{inel}}\). At 4~K, phonon absorption is
strongly suppressed; emission is only available once carriers have already
been accelerated above the relevant phonon energies. Consequently,
\(\nu_{\mathrm{inel}}(\varepsilon,T)\) is reduced over much of the
subthreshold energy range, lengthening inelastic free flights and increasing
the probability of reaching \(\varepsilon_i\)---a key mechanism lowering the
PI-predicted \(E_{\mathrm{crit}}\) at cryogenic temperatures.

Elastic, momentum-randomizing processes (ionized-impurity Brooks--Herring and
neutral-impurity scattering~\cite{BrooksHerring}) strongly affect drift
velocity and the low-field mobility but do not directly remove energy. They
enter through calibration of the total scattering content via the measured
mobility,
\begin{equation}
\mu(T)=\frac{q}{m_0^\ast}
\left\langle \frac{1}{\nu_{\mathrm{el}}(\varepsilon,T)+\nu_{\mathrm{inel}}(\varepsilon,T)}\right\rangle_{\varepsilon\sim k_B T}\!,
\label{eq:mu}
\end{equation}
which we use to constrain prefactors in
\(\nu_{\mathrm{ac}},\nu_{\mathrm{op}},\nu_{\mathrm{iv}},\nu_{\mathrm{def}}\).
At 4~K, the suppression of absorption increases
\(\tau_{\mathrm{inel}}=1/\nu_{\mathrm{inel}}\) and extends high-energy flights,
providing a direct physical basis for the strong temperature dependence of
avalanche onset~\cite{Jacoboni1983}.

\subsection{Lucky-drift tail and the ionization coefficient}

Under a field \(E\), the energy gain per unit length is
\(d\varepsilon/dx\simeq qE\). The probability that a carrier
reaches the ionization threshold \(\varepsilon_i\) without undergoing an
inelastic energy-loss event---the ``lucky drift'' probability---is
given by~\cite{Shockley1961,Grant1973}
\begin{equation}
P_{\mathrm{lucky}}(E)=\exp\!\left[-\frac{1}{qE}\int_0^{\varepsilon_i}
\frac{d\varepsilon}{v(\varepsilon)\,\tau_{\mathrm{inel}}(\varepsilon,T)}\right]
\label{eq:Plucky}
\end{equation}
with \(\tau_{\mathrm{inel}}(\varepsilon,T)=1/\nu_{\mathrm{inel}}(\varepsilon,T)\).

Once the carrier energy exceeds threshold, impact ionization proceeds with an
effective post-threshold mean free path \(\lambda_{\mathrm{eff}}\), leading to
the ionization coefficient
\begin{equation}
    \alpha(E,T) \approx \frac{P_{\text{lucky}}(E)}{\lambda_{\text{eff}}} \simeq A(T) \exp\left[-\frac{B(T)}{E}\right].
    \label{eq:alpha}
\end{equation}
The temperature-dependent coefficients are
\begin{equation}
    B(T) = \frac{1}{q} \int_0^{\varepsilon_i} \frac{d\varepsilon}{v(\varepsilon) \tau_{\text{inel}}(\varepsilon,T)}
    \quad \text{and} \quad
    A(T) \sim \frac{1}{\lambda_{\text{eff}}},
    \label{eq:AB}
\end{equation}
recovering the classic Chynoweth/Okuto--Crowell form widely observed in
semiconductors~\cite{Chynoweth1958,OkutoCrowell1974}. Here \(A(T)\) has units
of \(\mathrm{cm^{-1}}\) and \(B(T)\) has units of \(\mathrm{V/cm}\). In the
parabolic, constant-\(\tau\) limit, \(B\to \sqrt{2\varepsilon_i/m^\ast}/\mu\),
so the SFF scale emerges as the limiting case of the transport integral.

\subsection{Practical parameterization for Ge (77~K, 4~K)}

For device design, it is useful to (i) retain the physics in \(B(T)\) through
Kane \(v(\varepsilon)\) and a calibrated \(\tau_{\mathrm{inel}}(\varepsilon,T)\),
while (ii) treating \(A(T)\) as a weakly field-dependent prefactor tied to an
effective post-threshold path length \(\lambda_{\mathrm{eff}}\) (typically
nm--tens of nm near threshold). A practical workflow is:
\begin{enumerate}
\item Fix \(m_0^\ast\) (electrons in L valleys; holes: heavy/light admixture) and
choose a nominal \(\alpha_{\mathrm K}\) range; calibrate
\(\nu_{\mathrm{ac}},\nu_{\mathrm{op}},\nu_{\mathrm{iv}},\nu_{\mathrm{def}}\) so
Eq.~\eqref{eq:mu} matches measured \(\mu(T)\) at 77~K and 4~K~\cite{Jacoboni1983,Conwell1967,SzeNg}.
\item Compute \(B(T)\) from Eq.~\eqref{eq:AB} using \(\varepsilon_i\) in the range
\([E_g,\,W]\) (Ge: \(E_g\simeq 0.73\) eV at 77~K; \(W\simeq 2.96\) eV at cryogenic \(T\)),
and tabulate \(B\) for design use~\cite{SzeNg}.
\item Extract \(A(T)\) either from short-diode measurements (Sec.~\ref{sec:calibration})
or from calibrated Monte Carlo fits~\cite{OkutoCrowell1974,Jacoboni1983}.
\end{enumerate}

\subsection{Electron--hole asymmetry and excess noise}
Impact ionization in Ge is intrinsically asymmetric: the hole and electron
ionization coefficients, $\alpha(E,T)$ and $\beta(E,T)$, generally differ and
in the relevant high-field regime Ge often exhibits $\alpha>\beta$
(hole ionization larger than electron ionization)~\cite{Miller1955,Mikawa1980,SzeNg}.
This asymmetry governs both the mean multiplication and the excess noise.

Throughout this work we define $\alpha(E,T)$ as the hole ionization
coefficient and $\beta(E,T)$ as the electron ionization coefficient.  To express
the degree of \emph{near-unipolar} multiplication with a parameter bounded by
unity, we adopt the ratio of the smaller to the larger coefficient,
\begin{equation}
k \;\equiv\; \frac{\beta}{\alpha}\;\le\;1,
\label{eq:kdef}
\end{equation}
which is the appropriate asymmetry parameter for \emph{hole-initiated}
multiplication when $\alpha>\beta$. With this convention, $k\ll 1$
corresponds to strongly hole-dominated ionization (near-unipolar gain), the
regime that minimizes excess noise. Note that our $\alpha$ corresponds to $\beta$ in McIntyre’s original notation~\cite{McIntyre1966}.

For a hole-initiated geometry, the mean multiplication is
\begin{equation}
\begin{aligned}
M_h \;=\; \exp\!\left(\int_{0}^{d}\alpha\!\big(E(x),T\big)\,dx\right)
\qquad \\
\xrightarrow[\text{uniform }E]{}\qquad
M_h \;=\; \exp\!\big(\alpha\,d\big)\,,
\end{aligned}
\label{eq:mean-gain}
\end{equation}

where $d$ is the effective multiplication width. The excess noise factor $F$
quantifies the stochastic broadening introduced by the avalanche process
relative to shot noise at the same mean gain~\cite{McIntyre1966,SzeNg,OkutoCrowell1974}.
For hole-initiated multiplication with the convention in
Eq.~\eqref{eq:kdef}, the McIntyre expression may be written as
\begin{equation}
F_h \;\approx\; k\,M_h \;+\; \left(2-\frac{1}{M_h}\right)(1-k).
\label{eq:excess-noise}
\end{equation}
Equation~\eqref{eq:excess-noise} makes the design implication explicit: when
$k\ll 1$ (i.e., $\beta\ll\alpha$), the excess noise approaches the
near-unipolar limit \(F_h \rightarrow 2-1/M_h\), which is the minimum-noise
behavior for a given gain. Conversely, operating in an electron-initiated
configuration in p-type Ge when $\beta>\alpha$ promotes stronger bipolar feedback and a
rapidly increasing noise factor. Therefore, low-noise p-type Ge ICA architectures
should favor hole-initiated multiplication through appropriate contact
selection and field shaping~\cite{SzeNg,OkutoCrowell1974}.

\subsection{Calibration from measurements: \(M(V)\) and Chynoweth plots}

A robust extraction route proceeds via short, uniform-field test diodes:
\begin{enumerate}
\item Measure multiplication \(M(V)\) at fixed \(T\) and known \(d\). For
unipolar gain, \(\alpha=\ln M/d\) (or \(\beta=\ln M/d\) for electron initiation).
\item Plot \(\ln\alpha\) versus \(1/E\) (Chynoweth plot). Linear regions yield
slope \(B(T)\) and intercept \(\ln A(T)\)~\cite{Chynoweth1958,OkutoCrowell1974}.
\item Repeat at 77~K and 4~K to obtain \(A(T),B(T)\), then apply
Eq.~\eqref{eq:EcritPI} to predict \(E_{\rm crit}\) for target geometries.
\end{enumerate}
This workflow uses the SFF estimate as an upper-bound cross-check (via
\(\sqrt{2\varepsilon_i/m^\ast}/\mu\)) while anchoring the transport-informed
design in directly measured \((A,B)\).

\subsection{Additional real-world modifiers}

Space charge (trap filling), surface leakage/dead layers (a-Ge vs.\ Li-diffused
contacts), and self-heating modify both the field profile \(E(x)\) and the
effective \(\alpha(E)\), generally reducing the usable field relative to
transport-only predictions~\cite{SzeNg}. Capturing these effects requires
technology computer-aided design (TCAD), i.e., self-consistent
Poisson--drift--diffusion simulations augmented with calibrated
impact-ionization laws \(\alpha(E,T)\), leakage paths, and trap/recombination
kinetics. In practice, one solves
\begin{align}
\nabla\!\cdot\!\bigl(\epsilon \nabla \phi\bigr)
&= -q\bigl(p-n+N_D^+ - N_A^-\bigr), \label{eq:poisson}\\
\mathbf{J}_{n} &= q\mu_{n} n\,\mathbf{E} + q D_{n}\nabla n, \qquad
\mathbf{J}_{p} = q\mu_{p} p\,\mathbf{E} - q D_{p}\nabla p, \label{eq:drift-diff}
\end{align}
with recombination-generation terms that include Shockley--Read--Hall (SRH)
traps~\cite{ShockleyRead1952,Hall1952}, surface-recombination velocities,
series resistances, and (when needed) an electrothermal equation for
self-heating.

\paragraph*{Parameter definitions.}
In Eqs.~\eqref{eq:poisson}--\eqref{eq:drift-diff}, \(\epsilon\) is the
permittivity, \(\phi\) the electrostatic potential, \(q\) the elementary
charge, \(n\) and \(p\) the electron and hole densities, \(N_D^+\) and \(N_A^-\)
the ionized donor/acceptor densities, and \(\mathbf{E}=-\nabla\phi\) the
electric field. The mobilities \(\mu_{n,p}\) may be field- and
temperature-dependent, and \(D_{n,p}\) are diffusion coefficients (linked to
\(\mu_{n,p}\) by Einstein relations at the local lattice temperature). For SRH
recombination, a single-level trap at energy \(E_t\) with density \(N_t\) gives
\[
U_{\mathrm{SRH}}
=\frac{np-n_i^2}{\tau_p\bigl(n+n_1\bigr)+\tau_n\bigl(p+p_1\bigr)}\,,
\]
where \(n_1=n_i \exp\!\bigl((E_t-E_i)/k_BT\bigr)\),
\(p_1=n_i \exp\!\bigl((E_i-E_t)/k_BT\bigr)\), \(E_i\) is the intrinsic level,
and \(\tau_{n,p}=(\sigma_{n,p} v_{\mathrm{th}} N_t)^{-1}\) are lifetimes set by
capture cross sections \(\sigma_{n,p}\) and thermal velocity
\(v_{\mathrm{th}}\)~\cite{ShockleyRead1952,Hall1952}. Surface leakage is
modeled via boundary conditions with surface-recombination velocities
\(S_{n,p}\) and, where appropriate, fixed-charge or work-function offsets to
represent a-Ge vs.\ Li-diffused contacts. Self-heating may be included through
\[
\nabla\!\cdot\!\bigl(\kappa \nabla T\bigr) + H = 0,
\]
with thermal conductivity \(\kappa\), temperature \(T\), and volumetric heating
\(H\) (e.g., Joule heating \(=\mathbf{J}\!\cdot\!\mathbf{E}\)), enabling
electrothermal feedback analysis under high field.

\paragraph*{Usage in design.}
Equation~\eqref{eq:alpha} supplies the central material parameterization of
impact ionization; TCAD maps \(\alpha(E,T)\) onto the actual, possibly
nonuniform field \(E(x)\) to predict onset, identify hot spots, quantify gain
uniformity, and estimate excess-noise trends in geometries with realistic
contacts and surfaces.

\subsection{Design figure: \(E_{\rm crit}\) vs mobility}

To visualize design leverage, Fig.~\ref{fig:Ecrit_vs_mu} plots the SFF
upper-bound critical field versus mobility over
\(2\times10^3\)--\(2\times10^7\)~cm\(^2\)/V$\cdot$s for electrons
\((m^\ast=0.12\,m_0)\) and holes \((m^\ast=0.28\,m_0)\), comparing
\(\varepsilon_i=2.96\)~eV and \(0.73\)~eV. The \(E_{\rm crit}\propto \mu^{-1}\)
and \(\propto \sqrt{\varepsilon_i}\) scalings appear as parallel, vertically
shifted families of curves, with hole branches above electron branches due to
the larger effective mass. These trends provide quick targets for uniform-field
regions, while the PI transport model above supplies the physically grounded
reduction from the SFF upper bound through the integral \(B(T)\) in
Eq.~\eqref{eq:AB}.

\section{Device-level breakdown criterion}
\label{sec:device_level}

The transport model in Sec.~\ref{sec:physics-informed} provides a compact
parameterization of the ionization coefficient \(\alpha(E,T)\) (and \(\beta(E,T)\))
that captures the high-energy ``lucky drift'' tail and its temperature dependence.
To translate these \emph{material-level} laws into \emph{device-level} design
targets, we now connect \(\alpha(E,T)\) to a geometric multiplication width \(d\)
and formulate a practical onset (breakdown) criterion in terms of the field and
bias.

\medskip
\noindent
For a uniform-field multiplication region of width \(d\), avalanche onset is
reached when the \emph{mean} number of impact events accumulated across the
region is of order unity. For a \emph{unipolar} (hole-initiated) structure
this leads to the Townsend-type criterion
\begin{equation}
\int_0^d \alpha\!\big(E,T\big)\,dx \;\approx\; 1
\Longrightarrow
\alpha\!\big(E_{\mathrm{crit}},T\big)\,d \;\simeq\; 1 \quad (\text{uniform }E), 
\label{eq:townsend}
\end{equation}
where \(\alpha(E,T)\) is the hole ionization coefficient (units
\(\mathrm{cm^{-1}}\))~\cite{SzeNg,OkutoCrowell1974,McIntyre1966}. Using the
transport-informed Chynoweth/Okuto--Crowell form
\(\alpha(E,T)\simeq A(T)\,e^{-B(T)/E}\) gives the closed-form
\begin{equation}
\boxed{E_{\mathrm{crit}}^{\mathrm{(PI)}}(T,d)=\frac{B(T)}{\ln\!\big[A(T)\,d\big]}}\,, 
\label{eq:EcritPI}
\end{equation}
which makes explicit how transport (\(A,B\)) and geometry (\(d\)) jointly set
the onset field. In this picture, the breakdown (or onset) voltage follows
directly as
\begin{equation}
V_{\mathrm{B}}^{\mathrm{(PI)}}(T,d)
\;=\; E_{\mathrm{crit}}^{\mathrm{(PI)}}\,d
\;=\; d\,\frac{B(T)}{\ln\!\big[A(T)\,d\big]}\,.
\label{eq:Vb}
\end{equation}

\paragraph*{Assumptions and scope.}
Equations~\eqref{eq:EcritPI}--\eqref{eq:Vb} assume: (i) a \emph{uniform} field
across the multiplication width; (ii) unipolar multiplication (one carrier
species dominates, e.g., holes with \(\beta\!\ll\!\alpha\)); and (iii) a
locally exponential \(\alpha(E)\) parameterization extracted from transport or
measurement~\cite{OkutoCrowell1974,Chynoweth1958,Jacoboni1983}. These conditions
are well approximated in short planar test diodes and can be approached in
practical detectors via field shaping and guard structures.

\paragraph*{Design sensitivities and scaling laws.}
Equation~\eqref{eq:EcritPI} highlights several useful scalings:
\begin{itemize}
\item \textbf{Thickness leverage is logarithmic.} Differentiating gives
\(\partial E_{\mathrm{crit}}/\partial d
= -\,B\big/\!\big[d\,(\ln(A d))^2\big]\): increasing \(d\) lowers
\(E_{\mathrm{crit}}\), but with diminishing returns (\(\propto 1/\ln d\)).
Consequently, \(V_{\mathrm{B}} \propto d/\ln d\) grows \emph{sublinearly} with
thickness.
\item \textbf{Temperature/material leverage enters through \(A,B\).} Lower \(T\)
suppresses inelastic phonon channels and reduces the energy-relaxation integral
\(B(T)\) (often with a modest increase in \(A(T)\)), thereby lowering
\(E_{\mathrm{crit}}\) at fixed \(d\)~\cite{Jacoboni1983}. Improved purity
reduces defect-assisted inelastic pathways, also tending to reduce \(B\).
\item \textbf{Threshold energy.} Since \(B\) integrates energy relaxation up to
\(\varepsilon_i\), larger microscopic thresholds increase \(B\) and raise
\(E_{\mathrm{crit}}\) (cf.\ Sec.~\ref{sec:physics-informed}).
\end{itemize}

\paragraph*{Extraction of \(A(T)\) and \(B(T)\) from data.}
In a short, uniform-field diode of known \(d\), the multiplication factor
\(M(V)\) yields \(\alpha = (\ln M)/d\) for unipolar gain. Plotting \(\ln \alpha\)
versus \(1/E\) (the Chynoweth plot) produces a quasi-linear segment whose
slope/intercept give \(B(T)\) and \(\ln A(T)\), respectively~\cite{Chynoweth1958,OkutoCrowell1974,SzeNg}. Repeating at 77~K and 4~K provides the
calibrated \(A(T),B(T)\) pairs required in Eq.~\eqref{eq:EcritPI} for
geometry-to-bias translation.

\paragraph*{Bipolar generalization and noise.}
When both carriers ionize (\(\alpha,\beta\neq 0\)), the onset condition is
stricter than Eq.~\eqref{eq:townsend}; the coupled Townsend equations admit
solutions only below a critical field determined jointly by \(\alpha(E)\) and
\(\beta(E)\), while the mean gain and the \emph{excess noise factor} \(F\) are
governed by their ratio \(k=\beta/\alpha\)~\cite{McIntyre1966,SzeNg}. Although a
simple closed form for \(E_{\mathrm{crit}}\) analogous to
Eq.~\eqref{eq:EcritPI} is generally not available, the message for design is
clear: suppressing feedback from the non-injected carrier (small effective
\(k\)) both \emph{reduces} excess noise and \emph{relaxes} onset requirements
for a given \(d\).

\paragraph*{Nonuniform fields and hot-spot dominance.}
For spatially varying fields, the onset condition becomes
\(\int_0^d A\,e^{-B/E(x)}dx \simeq 1\).
Because the integrand is a rapidly varying function of \(1/E\), the highest
local-field regions dominate the integral. In practice, edge curvature,
guard-ring geometry, and sidewall roughness can create local hot-spots that
trigger microplasma breakdown before the mean field reaches the uniform-field
prediction of Eq.~\eqref{eq:EcritPI}~\cite{SzeNg}. This motivates the combined
workflow emphasized in Sec.~\ref{sec:physics-informed}: calibrate
\(\alpha(E,T)\) on uniform-field structures, then map it onto realistic
\(E(x)\) profiles using TCAD to identify and mitigate hot spots.

\noindent\textit{Effective-slab approximation for a nonuniform point-contact field.}
In the point-contact geometry the electric field is highly nonuniform, \(E(\mathbf{r})\), and the avalanche condition is governed by the path integral \(\int \alpha[E(x),T]\,dx\). For the analytic estimates and the illustrative \(M(V)\) curves in this work, we approximate the localized high-field region beneath the anode by an \emph{effective uniform slab} of thickness \(d_{\mathrm{eff}}\) with a representative field \(E_{\mathrm{eff}}\), chosen so that \(\int \alpha[E(x),T]\,dx \approx \alpha(E_{\mathrm{eff}},T)\,d_{\mathrm{eff}}\) (equivalently, \(d_{\mathrm{eff}}\) may be taken as the extent of the region where \(E\) remains near its peak in TCAD). This mapping captures the hot-spot dominance of \(\alpha\), but it can fail when sub-dominant geometric features (edge curvature, surface asperities, guard-ring corners) create localized microplasma-prone hot spots that satisfy the avalanche criterion earlier than the mean-field estimate. These failure modes motivate using TCAD-derived \(E(\mathbf{r})\) together with diode \(M(V)\) calibration to bound \(d_{\mathrm{eff}}\), identify hot spots, and determine the device-limited onset bias.

\paragraph*{Operational definition of $E_{\mathrm{crit}}$ (uniform-field vs.\ device-limited onset).}
In this work, we define the \emph{operational} onset field $E_{\mathrm{crit}}$ as the field at which internal charge amplification becomes \emph{measurably} and \emph{stably} observable under controlled conditions. For a uniform-field test diode, $E_{\mathrm{crit}}^{(\mathrm{PI})}(T,d)$ denotes the \emph{uniform-field onset} predicted by the Townsend criterion $\int_0^d \alpha(E,T)\,dx \simeq 1$ (Eq.~\eqref{eq:townsend}), equivalently the point where the measured multiplication $M(V)$ departs from unity beyond the experimental noise floor and remains reproducible under repeated bias sweeps and fixed-$T$ holds (no runaway leakage or irreversible conditioning). In contrast, practical Ge ICA geometries generally exhibit spatially varying $E(x)$; in that case the observed onset is often \emph{device-limited} and can occur at a lower \emph{applied bias} than the uniform-field prediction when localized hot-spots (edge curvature, surface roughness, or microplasma sites) satisfy the avalanche condition first. We therefore distinguish the \emph{material/onset scale} set by $E_{\mathrm{crit}}^{(\mathrm{PI})}$ (transport + effective multiplication width) from the \emph{device-limited onset}, which is governed by the maximum local field and surface/space-charge feedback and must be assessed with calibrated TCAD field maps and leakage/conditioning data.

\paragraph*{Illustrative numerics (order-of-magnitude).}
Using the Chynoweth form \(\alpha(E,T)=A(T)\exp[-B(T)/E]\), a representative
post-threshold inelastic path length \(\Lambda_i^{\rm eff}\sim 20\)–\(50~\mathrm{nm}\)
implies
\[
A(T)\sim \frac{1}{\Lambda_i^{\rm eff}} \approx (2\text{–}5)\times10^{5}~\mathrm{cm^{-1}},
\]
with only a weak temperature dependence expected unless \(\Lambda_i^{\rm eff}\)
is independently measured.
Transport fits at 77~K typically give
\(B(77~\mathrm{K})\sim(6\text{–}9)\times10^{4}~\mathrm{V/cm}\), consistent with
Ge-scale ionization integrals~\cite{OkutoCrowell1974,Jacoboni1983}. For a
representative gap \(d=10~\mu\mathrm{m}\) (\(10^{-3}~\mathrm{cm}\)) we obtain
\(A(77~\mathrm{K})\,d\sim 200\text{–}500\) and \(\ln[A(77~\mathrm{K})\,d]\sim
5.3\text{–}6.2\), yielding \(E_{\mathrm{crit}}\sim 10\text{–}17~\mathrm{kV/cm}\).

To connect these estimates to cryogenic operation without introducing ad hoc
parameters, we follow the temperature scaling implied by the PI/SFF limit:
under the same transport assumptions, the ionization exponent inherits the
mobility dependence such that \(B(T)\propto 1/\mu(T)\) (see
Sec.~\ref{sec:physics-informed}). Thus, adopting the commonly used scaling
\(\mu(4~\mathrm{K})\approx 10\,\mu(77~\mathrm{K})\) gives
\[
B(4~\mathrm{K}) \approx B(77~\mathrm{K})\,\frac{\mu(77~\mathrm{K})}{\mu(4~\mathrm{K})}
\approx \frac{B(77~\mathrm{K})}{10},
\]
while \(A(T)\) is taken to be approximately constant to first order,
\(A(4~\mathrm{K})\approx A(77~\mathrm{K})\), in the absence of a calibrated
\(\Lambda_i^{\rm eff}(T)\). With this mobility-driven reduction in \(B\),
\(E_{\mathrm{crit}}\) decreases at 4~K, consistent with suppressed inelastic
relaxation at cryogenic temperature~\cite{Jacoboni1983}. These ranges provide
practical design targets prior to derating for field nonuniformities and
surface effects.

\medskip
To move beyond back-of-the-envelope estimates and provide design-ready
numerics, we include one explicit \emph{illustrative} parameter set constructed
from the above temperature dependence. Table~\ref{tab:AB_examples} lists
representative Chynoweth coefficients \((A_h(T),B_h(T))\) and \((A_e(T),B_e(T))\)
for holes and electrons at 77~K and 4~K. The 77~K values are chosen within the
order-of-magnitude ranges quoted above, and the 4~K values are obtained by
applying \(B(T)\propto 1/\mu(T)\) with \(\mu(4~\mathrm{K})/\mu(77~\mathrm{K})=10\)
and taking \(A(4~\mathrm{K})\approx A(77~\mathrm{K})\). These coefficients
reproduce the example \(\alpha(E)\) curves used in Figs.~\ref{fig:chynoweth} and
\ref{fig:MV_curve}, but should be regarded as \emph{model-based design examples}
rather than calibrated experimental values. Ultimately, precise \((A(T),B(T))\)
pairs should be extracted from dedicated measurements on short Ge test
structures over temperature and field.

\begin{table}[t]
  \centering
  \caption{Illustrative Chynoweth parameters \((A(T),B(T))\) for electrons and holes
  in high-purity Ge at 77~K and 4~K. The 77~K values are selected within the
  order-of-magnitude ranges discussed in the text. The 4~K values are obtained
  by applying the temperature dependence implied by the PI/SFF limit,
  \(B(T)\propto 1/\mu(T)\), using \(\mu(4~\mathrm{K})/\mu(77~\mathrm{K})=10\),
  and taking \(A(4~\mathrm{K})\approx A(77~\mathrm{K})\) in the absence of
  calibrated \(\Lambda_i^{\rm eff}(T)\). These are model-consistent design
  examples, not experimentally calibrated coefficients.}
  \label{tab:AB_examples}
  \begin{tabular}{lccc}
    \hline\hline
    Carrier type & Temperature & \(A\) (cm\(^{-1}\)) & \(B\) (V/cm) \\
    \hline
    Holes & 77 K & \(3.0\times10^{5}\) & \(7.5\times10^{4}\) \\
    Holes & 4 K  & \(3.0\times10^{5}\) & \(7.5\times10^{3}\) \\
    Electrons     & 77 K & \(2.5\times10^{5}\) & \(8.5\times10^{4}\) \\
    Electrons     & 4 K  & \(2.5\times10^{5}\) & \(8.5\times10^{3}\) \\
    \hline\hline
  \end{tabular}
\end{table}

\medskip
In summary, Eqs.~\eqref{eq:EcritPI}--\eqref{eq:Vb} provide a compact,
physics-informed bridge between microscopic transport (\(A(T),B(T)\)) and device
geometry (\(d\)), enabling rapid first-pass estimates of avalanche onset and
breakdown voltage that can then be refined with calibrated TCAD and \(M(V)\)
measurements.

\section{Bridging SFF and the PI model}
\label{sec:device_design}

The device-level criterion derived in Sec.~\ref{sec:device_level} converts a
material law \(\alpha(E,T)\) into an onset field through the finite-width
condition \(\int_0^d \alpha\,dx \simeq 1\). What remains is to connect this
physics-informed (PI) description back to the simpler single--free-flight (SFF)
estimate of Sec.~\ref{sec:sff}, which is often used as a first-pass design
target. In this section we make that bridge explicit, clarify the role of the
temperature-dependent Chynoweth parameter \(B(T)\) (especially at 4~K), and
summarize a practical calibration path for \((A(T),B(T))\).

\paragraph*{Asymptotic reduction to SFF (and its temperature scaling).}
The SFF result emerges as the parabolic, constant-\(\tau\) limit of the PI
energy-relaxation integral. If one approximates the carrier velocity by the
parabolic-band form \(v(\varepsilon)\!\approx\!\sqrt{2\varepsilon/m^\ast}\) and
treats the inelastic time as energy-independent up to threshold,
\(\tau_{\mathrm{inel}}(\varepsilon,T)\!\approx\!\tau(T)\), then the PI
coefficient \(B(T)\) in \cref{eq:AB} reduces to
\begin{equation}
B(T)\;\approx\;\frac{1}{q}\int_0^{\varepsilon_i}
\frac{d\varepsilon}{\sqrt{2\varepsilon/m^\ast}\,\tau(T)}
=\frac{\sqrt{2m^\ast\varepsilon_i}}{q\,\tau(T)}
=\frac{\sqrt{2\varepsilon_i/m^\ast}}{\mu(T)}\,,
\end{equation}
where we used \(\mu(T)=q\tau(T)/m^\ast\). Substituting this \(B(T)\) into the
device rule \cref{eq:EcritPI} and taking \(\ln(A d)\!\to\!1\) recovers the
familiar SFF scaling \cref{eq:SFF}. Thus, SFF is the \emph{parabolic,
constant-\(\tau\) asymptote} of the PI framework~\cite{SzeNg}.

This expression also makes the cryogenic trend explicit: in the SFF asymptote,
\(B(T)\propto 1/\mu(T)\). Therefore, when the mobility increases strongly at
4~K, one expects \(B(4~\mathrm{K})\ll B(77~\mathrm{K})\). Since
\(\alpha(E,T)=A(T)\exp[-B(T)/E]\), a smaller \(B(T)\) implies a larger
\(\alpha\) at fixed field and, correspondingly, a lower onset field at 4~K.

\paragraph*{Corrections from nonparabolicity and energy-dependent scattering.}
In realistic Ge, two effects modify the constant-\(\tau\) SFF scaling. First,
at elevated carrier energies the Kane dispersion
\(
  \varepsilon(1+\alpha_{\mathrm K}\varepsilon)=\hbar^2k^2/(2m_0^\ast)
\)
reduces the group velocity \(v(\varepsilon)\) relative to the parabolic case.
Second, inelastic rates \(\nu_{\mathrm{inel}}(\varepsilon,T)\) grow with
\(\varepsilon\) as optical and intervalley phonon channels open.%
~\cite{Kane1957,Conwell1967,Jacoboni1983}
Both effects increase the integrand
\(
[v(\varepsilon)\tau_{\mathrm{inel}}(\varepsilon,T)]^{-1}
\)
and therefore increase the microscopic energy-relaxation integral
\[
  B(T)=\frac{1}{q}\int_0^{\varepsilon_i}
       \frac{d\varepsilon}{v(\varepsilon)\tau_{\mathrm{inel}}(\varepsilon,T)}\,,
\]
relative to the crude SFF evaluation that assumes a parabolic band and a
constant relaxation time. Taken in isolation at a \emph{fixed} temperature, a
larger \(B(T)\) shifts \(\alpha(E)=A\exp[-B/E]\) to higher fields.

At cryogenic temperature, however, the dominant trend relevant for avalanche
onset is that energy relaxation is suppressed: phonon absorption is strongly
reduced and inelastic relaxation times lengthen, i.e.
\(\tau_{\mathrm{inel}}(\varepsilon,4~\mathrm{K}) >
\tau_{\mathrm{inel}}(\varepsilon,77~\mathrm{K})\) over the onset-relevant energy
range. This increases the mean energy gained between inelastic events and
\emph{reduces} the energy-relaxation integral \(B(T)\) at 4~K relative to 77~K,
so that \(B(4~\mathrm{K})<B(77~\mathrm{K})\) is the physically expected
direction for Ge under otherwise similar conditions.~\cite{Jacoboni1983}

\paragraph*{Why the PI onset is often below the SFF reference (and when it need not be).}
The key conceptual distinction is that SFF encodes a \emph{single-flight
energy-gain} condition, whereas device onset is set by the \emph{finite-length}
criterion \(\alpha d\sim\mathcal{O}(1)\). In the Chynoweth representation this
yields
\[
E_{\mathrm{crit}}^{(\mathrm{PI})}(T,d) = \frac{B_{\mathrm{PI}}(T)}{\ln[A(T)\,d]}\,,
\]
\[
E_{\mathrm{crit}}^{(\mathrm{SFF})}(T)\sim B_{\mathrm{SFF}}(T)\quad(\ln(A d)\to 1).
\]
For typical Ge values \(A \sim 10^{5}\text{--}10^{7}~\mathrm{cm^{-1}}\) and
\(d \sim 5\text{--}50~\mu\mathrm{m}\), one has \(A d \sim 10\text{--}10^{3}\)
and hence \(\ln(A d)\sim 2\text{--}7\). Even when
\(B_{\mathrm{PI}}(T)\approx B_{\mathrm{SFF}}(T)\), this logarithmic factor
reduces the onset field by a factor of a few, yielding
\(E_{\mathrm{crit}}^{(\mathrm{PI})} < E_{\mathrm{crit}}^{(\mathrm{SFF})}\) as a
common outcome.

Importantly, this ordering is not a mathematical identity: if one uses
parameter sets in which \(B_{\mathrm{PI}}(T)\) is substantially larger than the
SFF-like estimate (e.g., due to inconsistent or non-physical temperature inputs
for \(A(T)\) and \(B(T)\)), then the computed
\(E_{\mathrm{crit}}^{(\mathrm{PI})}\) can exceed the SFF reference.
To avoid confusion, we therefore treat SFF as a \emph{one-flight reference
scale} (the parabolic constant-\(\tau\) asymptote), not a universal bound across
all parameterizations. In the temperature-consistent PI picture emphasized
here, the cryogenic trend \(B(4~\mathrm{K})<B(77~\mathrm{K})\) pushes
\(\alpha(E)\) upward and further lowers the onset field at low
temperature.~\cite{Jacoboni1983}

\begin{figure}[t]
  \centering
  \includegraphics[width=0.9\linewidth]{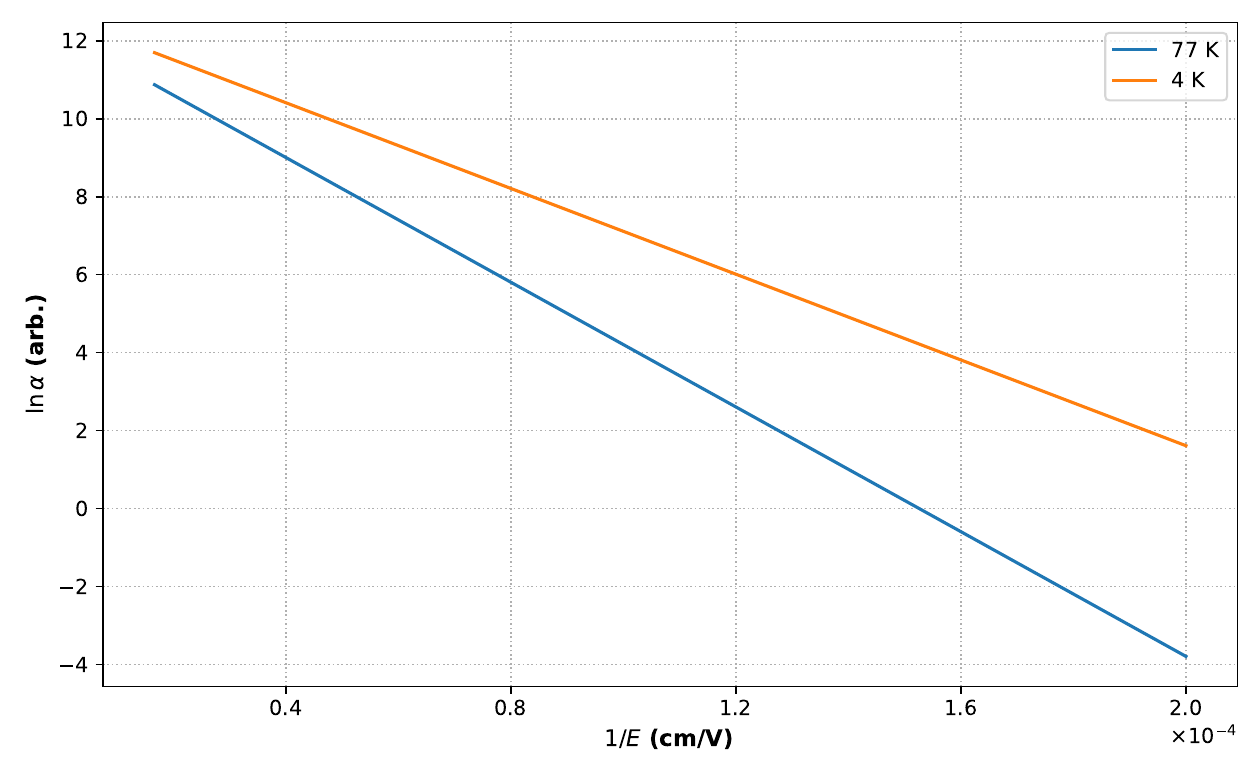}
  \caption{\textbf{Chynoweth plot (conceptual).} Illustration of \(\ln\alpha\) vs \(1/E\) for two temperatures using \(\alpha(E)=A\,e^{-B/E}\) with \(B(4\mathrm{K})<B(77\mathrm{K})\). The slope gives \(B(T)\) and the intercept \(\ln A(T)\), enabling direct calibration from \(M(V)\) data on uniform-field diodes~\cite{Chynoweth1958,OkutoCrowell1974}.}
  \label{fig:chynoweth}
\end{figure}

\paragraph*{Chynoweth representation and parameter identification.}
The PI form
\(
\alpha(E,T)\simeq A(T)\exp[-B(T)/E]
\)
implies that \(\ln\alpha\) plotted versus \(1/E\) is quasi-linear over the
onset-relevant field range, with slope \(B(T)\) and intercept \(\ln A(T)\).%
~\cite{Chynoweth1958,OkutoCrowell1974}
This provides a direct experimental route to extract \((A,B)\) from short,
uniform-field test diodes: measure \(M(V)\), compute \(\alpha=\ln M/d\), and
build the Chynoweth plot at each temperature. An illustrative example is shown
in \cref{fig:chynoweth}, emphasizing the smaller \(B\) at 4~K than at 77~K and
therefore a larger \(\alpha\) at the same \(E\).

\begin{figure}[t]
  \centering
  \includegraphics[width=0.9\linewidth]{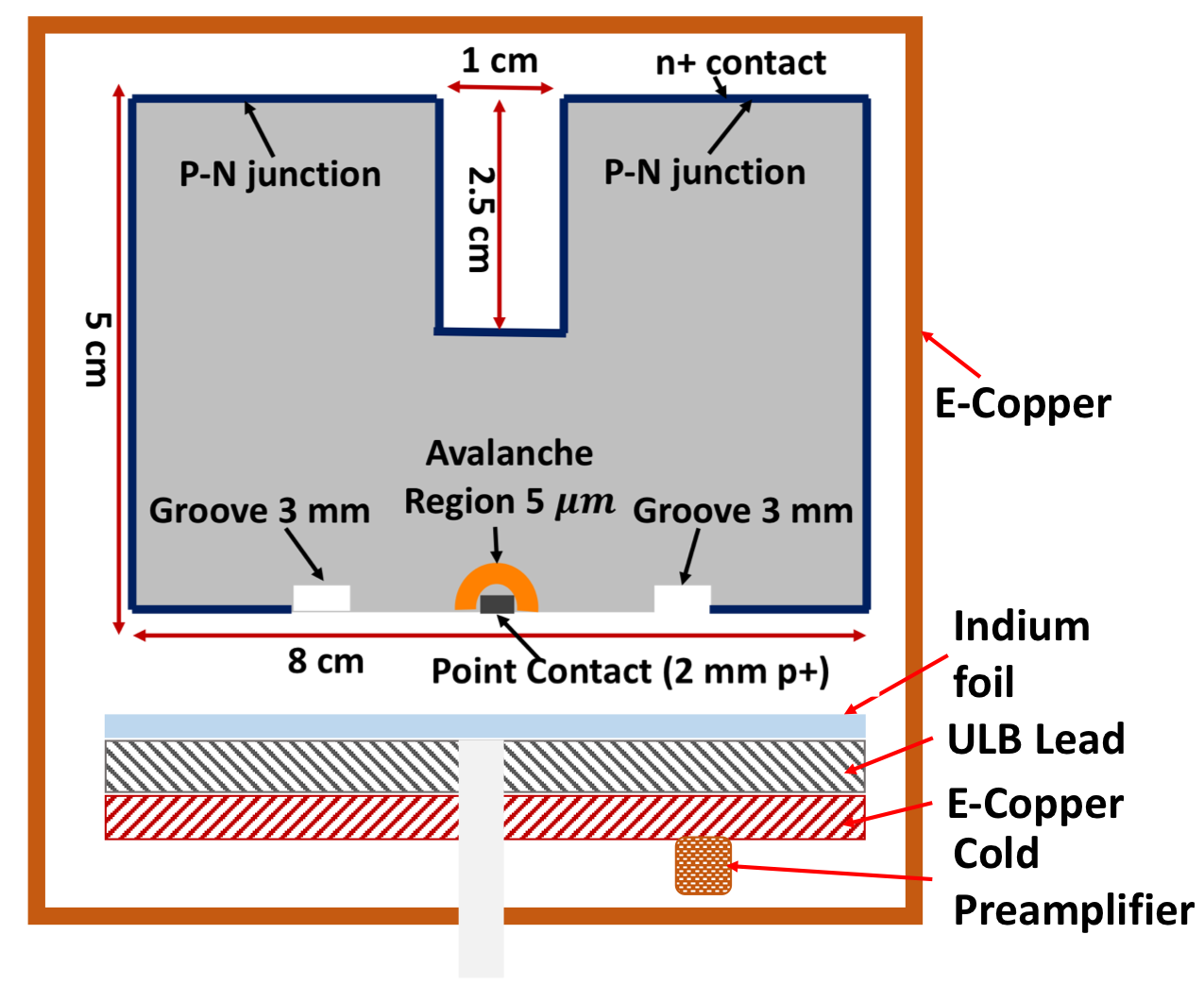}
  \caption{\textbf{Conceptual cross-section of a high-purity Ge internal-charge-amplification (ICA) detector.}
  A small p$^+$ point contact (2\,mm diameter) at the bottom defines a low-capacitance anode (target \(C\!\approx\!0.6\) pF), while an outer n$^+$ contact forms two bulk P--N junctions. A narrow, high-field \emph{avalanche region} (\(\sim 5~\mu\mathrm{m}\)) is engineered beneath the point contact to provide internal charge gain. Grooves (\(\sim 3~\mathrm{mm}\)) around the contact mitigate surface leakage and help shape the field. Illustrative operating parameters: leakage current \(<10~\mathrm{pA}\), depletion voltage \(\approx +3.2~\mathrm{kV}\), operating bias \(\approx +5.0~\mathrm{kV}\). The detector (footprint \(\sim 8~\mathrm{cm}\times 5~\mathrm{cm}\), central slot depth \(\sim 2.5~\mathrm{cm}\), top-contact separation \(\sim 1~\mathrm{cm}\)) is housed in electrolytic copper (E-Cu) with indium foil for thermal/mechanical coupling and surrounded by ultra-low-background (ULB) Pb shielding. A cold preamplifier is mounted close to the point contact to minimize stray capacitance and series noise. This geometry concentrates the field under the anode for stable ICA while maintaining low noise and excellent energy resolution.}
  \label{fig:ica_schematic}
\end{figure}

\paragraph*{Device connection and design figures.}
Once \((A,B)\) are calibrated, the uniform-field onset follows directly from
\cref{eq:EcritPI}:
\(
E_{\mathrm{crit}}^{(\mathrm{PI})}(T,d)=B(T)/\ln[A(T)\,d]\,.
\)
Because the dependence on \(d\) is only logarithmic, the dominant design levers
are: (i) temperature and purity/relaxation physics (primarily through \(B(T)\));
(ii) carrier asymmetry and injection control (minimize the effective
\(k\equiv\beta/\alpha\) to reduce excess noise); and (iii) field shaping that
concentrates gain where intended while suppressing hot-spots. A schematic
highlighting the multiplication width \(d\) and field localization is shown in
\cref{fig:ica_schematic}; the companion SFF scaling with \(\mu\) and
\(\varepsilon_i\) is visualized in \cref{fig:Ecrit_vs_mu}.

If one chooses a net impurity concentration in the range
\(5\times10^{9}\)–\(3\times10^{10}~\mathrm{cm^{-3}}\) for a p-type Ge ICA detector
geometry shown in \cref{fig:ica_schematic}, the depletion voltage at 77~K is
approximately \(3.2~\mathrm{kV}\), assuming that the p-type impurity
concentration varies along the crystal axis as
\[
C(z) = a + b z + n \exp\!\left(\frac{z - l}{m}\right),
\]
where \(a\), \(b\), \(n\), \(l\), and \(m\) are constants.
Using the Hall Effect measurements from many crystals grown in our lab, we
determined the parameters used in the simulations to be
\begin{align}
  a &= 4.61\times10^{9}~\mathrm{cm^{-3}}, &
  b &= -7.21\times10^{7}~\mathrm{cm^{-4}}, \nonumber\\
  n &= 2.25\times10^{10}~\mathrm{cm^{-3}}, &
  l &= -0.0878~\mathrm{cm}, \quad
  m = 0.722~\mathrm{cm},
\end{align}
with the detector coordinate origin at \(z_0 = 0~\mathrm{cm}\).
This parameterization yields net impurity concentrations of
\(\rho(0~\mathrm{cm}) \approx 3.0\times10^{10}~\mathrm{cm^{-3}}\) and
\(\rho(5~\mathrm{cm}) \approx 4.6\times10^{9}~\mathrm{cm^{-3}}\).

\begin{figure}[t]
  \centering
  \includegraphics[width=0.9\linewidth]{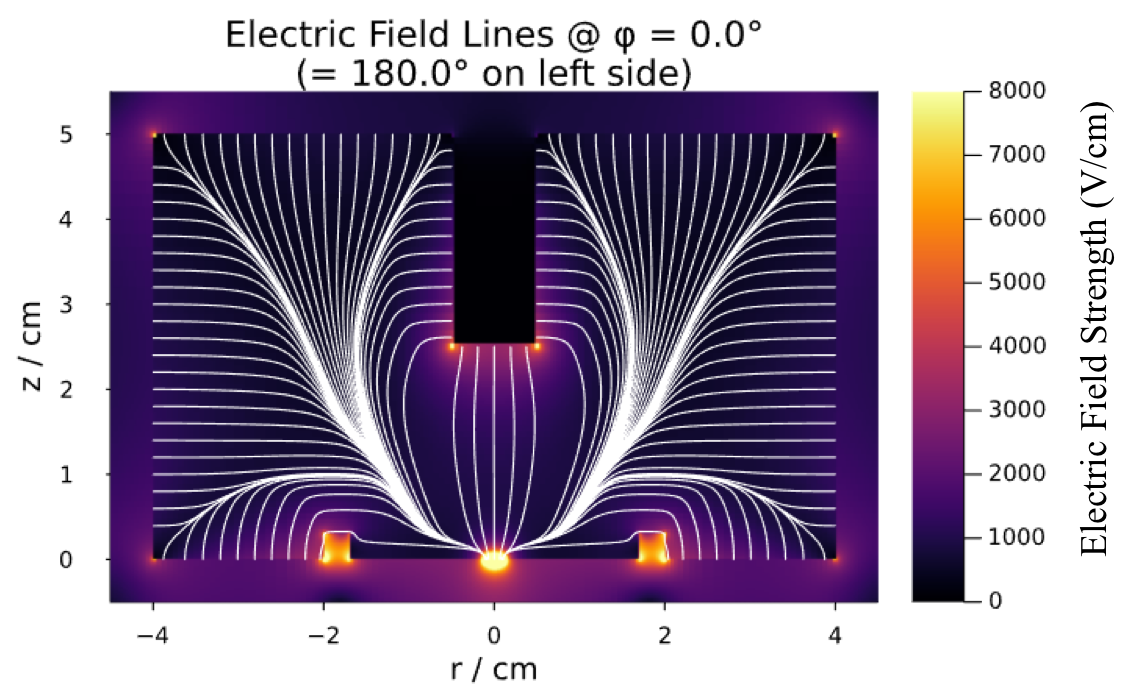}
  \caption{Shown is the electric-field distribution calculated using Julia~\cite{bezanson2017julia}. The highest field occurs in the region around the point contact, reaching values above 8000 V/cm.}
  \label{fig:field}
\end{figure}

When the detector is operated at \(5.0~\mathrm{kV}\), the resulting
electric-field distribution obtained from a TCAD simulation implemented
in Julia~\cite{Abt2021SSD,bezanson2017julia} is shown in
\cref{fig:field}.
As shown in Fig.~\ref{fig:field}, the peak electric field reaches
\(\sim 8\times 10^{3}~\mathrm{V/cm}\) in the vicinity of the point contact,
satisfying the critical-field condition for initiating the ICA process.
This field scale is consistent with the PI-predicted onset values for a
multiplication width of \(d \sim 5~\mu\mathrm{m}\) in Table~III.

To make this connection to the PI model explicit, we now evaluate
\(\,E_{\mathrm{crit}}^{(\mathrm{PI})}(T,d)\,\) for a parameter set consistent
with the Ge ICA geometry. Using the illustrative hole coefficients
of Table~\ref{tab:AB_examples}, for example
\(A_h \approx 3.0\times 10^{5}~\mathrm{cm^{-1}}\) and
\(B_h \approx 7.5\times 10^{4}~\mathrm{V/cm}\) at 77~K, and taking
\(d \simeq 5\)–\(8~\mu\mathrm{m}\) as the effective high-field
multiplication width from the TCAD field profile in Fig.~4, the
PI criterion
\(
E_{\mathrm{crit}}^{(\mathrm{PI})}
= B / \ln(A d)
\)
yields
\(E_{\mathrm{crit}}^{(\mathrm{PI})} \approx (7\text{–}9)\times 10^{3}~\mathrm{V/cm}\).
This range is in good agreement with the peak field
\(\sim 8\times 10^{3}~\mathrm{V/cm}\) obtained from the TCAD simulation at
\(5.0~\mathrm{kV}\), confirming that the Ge ICA operating bias is only
slightly above the PI-predicted avalanche onset. In the present Ge ICA
layout we assume hole-dominated multiplication: the outer
\(n^{+}\) contact is positively biased such that holes drift toward the small
\(p^{+}\) point contact through the high-field region, while electrons see a
much lower effective field. This corresponds to \(\beta(E) \ll \alpha(E)\)
and a small ionization-coefficient ratio \(k \equiv \beta/\alpha \ll 1\),
which, in the McIntyre expression
\(F(M) \approx k M + (2 - 1/M)(1-k)\),
keeps the excess noise factor in the range \(F \sim 2\)–\(3\) for
moderate gain \(M \sim 10\text{–}20\). Finally, using the PI-derived
\(\alpha(E,T)\) together with the TCAD field \(E(z)\), the expected
multiplication curve \(M_h(V)\approx \exp\!\left(\int_{0}^{d}\alpha_h\!\big(E(z),T\big)\,dz\right).
\)
rises from \(M \simeq 1\) just below \(5~\mathrm{kV}\) to
\(M \sim 10\)–\(20\) slightly above this bias for the Ge ICA geometry.
Even though these \(M(V)\) and \(F(M)\) estimates are based on model
parameters rather than experimental \(\alpha(E)\) calibration, they
demonstrate how the SFF\(\to\)PI framework can be carried through to
device-level observables for a realistic impurity gradient and TCAD
field distribution.

\paragraph*{Summary of the bridge.}
\begin{itemize}
\item \textbf{SFF \(\Rightarrow\) PI:} Replace \(v(\varepsilon)\!\to\!v_{\mathrm{Kane}}(\varepsilon)\) and \(\tau\!\to\!\tau_{\mathrm{inel}}(\varepsilon,T)\) inside the energy-relaxation integral for \(B(T)\); retain a weakly field-dependent \(A(T)\) tied to the post-threshold mean free path~\cite{Jacoboni1983,OkutoCrowell1974}.
\item \textbf{Experiment \(\Rightarrow\) \((A,B)\):} Use \(M(V)\) on short, uniform diodes to build \(\ln\alpha\) vs \(1/E\) (Chynoweth) and read off \((A,B)\) at each temperature~\cite{Chynoweth1958,OkutoCrowell1974}.
\item \textbf{Device onset:} Insert \((A,B)\) into \(\alpha d\simeq 1\) to obtain \(E_{\mathrm{crit}}^{(\mathrm{PI})}\), and compare to the SFF reference to quantify design margin. In a temperature-consistent model for Ge, \(B(4~\mathrm{K})<B(77~\mathrm{K})\) implies a lower onset field at 4~K.
\end{itemize}

\section{Calibration workflow and example numbers}
\label{sec:calibration}

Building on Sec.~\ref{sec:device_design}, where we connected the SFF bound to the
PI ionization law and to device-level observables, we now give a practical
calibration workflow. The goal is to extract \((A(T),B(T))\) from short,
uniform-field test structures and then use \cref{eq:EcritPI} (with TCAD
derating when needed) to predict onset fields and breakdown voltages for
realistic Ge ICA geometries. The key temperature-dependent update emphasized
here is that suppressed inelastic energy relaxation at \(4~\mathrm{K}\) yields
a \emph{smaller} Chynoweth slope parameter than at \(77~\mathrm{K}\),
i.e.\ \(B(4~\mathrm{K})<B(77~\mathrm{K})\), which lowers
\(E_{\mathrm{crit}}^{(\mathrm{PI})}\) and shifts multiplication to lower
bias for a fixed effective \(d\).~\cite{Jacoboni1983,OkutoCrowell1974}

\subsection{Calibration steps}

\paragraph*{Workflow (from transport to device).}
We outline a data-anchored procedure that turns measurements on short test diodes into
device-onset predictions:
\begin{enumerate}
  \item \textbf{Geometry \& metrology.} Fabricate uniform-field test structures of known multiplication width \(d\)
  (e.g., planar p\(^{+}\)–i–n slabs with a-Ge/Li contacts). Measure \(d\) by profilometry or process control;
  record series/contact resistances and surface conditions (passivation, bevel) for subsequent TCAD.
  \item \textbf{Low-field transport anchor.} Measure \(\mu(T)\) at 77~K (and 4~K) via Hall or time-of-flight;
  use \cref{eq:mu} to calibrate prefactors in \(\nu_{\mathrm{ac}},\nu_{\mathrm{op}},\nu_{\mathrm{iv}},\nu_{\mathrm{def}}\)
  so that the transport model reproduces the measured \(\mu(T)\)~\cite{SzeNg,Jacoboni1983,Conwell1967}.
  \item \textbf{High-energy kinematics.} Set the velocity–energy relation \(v(\varepsilon)\) using a Kane dispersion
  with a nominal \(\alpha_{\mathrm K}\) range for Ge (indirect-gap, L valleys), and band-edge masses \(m_0^\ast\)
  appropriate to electrons/holes~\cite{Kane1957,Jacoboni1983}.
  \item \textbf{Energy-relaxation integral.} Choose a physically motivated impact threshold \(\varepsilon_i\in[E_g,\,W]\)
  (Ge: \(E_g\!\approx\!0.73\) eV at cryo; pair-creation \(W\!\approx\!2.9\)–2.96 eV). Compute
  \[
    B(T)\;=\;\frac{1}{q}\int_0^{\varepsilon_i}\frac{d\varepsilon}{v(\varepsilon)\,\tau_{\mathrm{inel}}(\varepsilon,T)}
  \]
  using the calibrated \(\tau_{\mathrm{inel}}=1/\nu_{\mathrm{inel}}\) (acoustic, optical, intervalley, defect)~\cite{Jacoboni1983,Conwell1967}.
  The cryogenic expectation is that reduced phonon absorption at \(4~\mathrm{K}\)
  increases \(\tau_{\mathrm{inel}}\) over much of the relevant energy range,
  thereby reducing the integral and yielding \(B(4~\mathrm{K})<B(77~\mathrm{K})\).%
  ~\cite{Jacoboni1983}
  \item \textbf{Post-threshold prefactor.} Use \(A(T)\sim 1/\Lambda_i^{\mathrm{eff}}\) (tens of nm) as an initial scale,
  then \emph{refine} \(A,B\) by fitting measured multiplication \(M(V)\) on the same test diodes:
  \(\alpha=\ln M/d\), and a Chynoweth plot \(\ln\alpha\) vs \(1/E\) yields slope \(B(T)\) and intercept \(\ln A(T)\)~\cite{Chynoweth1958,OkutoCrowell1974}.
  \item \textbf{Device onset.} Insert \(\bigl(A(T),B(T)\bigr)\) into
  \(
  E_{\mathrm{crit}}^{(\mathrm{PI})}(T,d)=B(T)/\ln\!\big[A(T)d\big]
  \)
  (\cref{eq:EcritPI}) to predict onset field and \(V_{\mathrm{B}}=E_{\mathrm{crit}}d\) for a target geometry.
  Use TCAD to de-rate for nonuniform \(E(x)\) and surface/leakage effects (Sec.~\ref{sec:physics-informed}).
  \item \textbf{Uncertainty and temperature dependence.} Repeat at 4~K.
  Suppressed phonon absorption increases \(\tau_{\mathrm{inel}}\), reducing \(B(T)\) and thereby \(E_{\mathrm{crit}}\)~\cite{Jacoboni1983}.
  Propagate uncertainties from \(d\), \(\varepsilon_i\), and the \(\ln\)-fit confidence intervals of \(A,B\).
\end{enumerate}

The transport assumptions used to initialize the \((A,B)\) search space and to build
\(\nu_{\mathrm{inel}}(\varepsilon,T)\) are summarized in Table~\ref{tab:provenance}.
These ranges are narrowed during diode \(M(V)\) calibration (Sec.~\ref{sec:physics-informed})
to produce the final \((A,B)\) with confidence intervals.

\begin{table}[htp]
\centering
\caption{Transport assumptions used to illustrate the PI ionization model. Ranges reflect values commonly reported for Ge at cryogenic $T$; they were used to set the initial search space for $(A,B)$ prior to diode calibration.}
\label{tab:provenance}
\begin{tabular}{ll}
\toprule
\textbf{Quantity} & \textbf{Assumed range} \\
\midrule
Optical phonon energy $\hbar\omega_{\mathrm{op}}$ & $35$--$37$~meV~\cite{Jacoboni1983,Conwell1967} \\
Intervalley phonon energies $\hbar\Omega_{\mathrm{iv}}$ & $\sim 25$--$35$~meV~\cite{Jacoboni1983,Conwell1967} \\
Acoustic deformation potential (e) & $E_1 \sim 10$--$14$~eV~\cite{Jacoboni1983} \\
Ionized impurity density $N_{\mathrm{I}}$ & $10^9$--$10^{11}$~cm$^{-3}$\\
Neutral/defect centers $N_{\mathrm{D}}^{0}$ & $\lesssim 10^{10}$~cm$^{-3}$ \\
Kane nonparabolicity $\alpha_{\mathrm K}$ & $0.4$--$1.0$~eV$^{-1}$~\cite{Kane1957,Jacoboni1983} \\
Band-edge masses $(m^{\ast}_{e}, m^{\ast}_{h})$ & $(0.12,\,0.28)\,m_0$~\cite{SzeNg} \\
Pair-creation energy $W$ & $2.9$--$2.96$~eV~\cite{SzeNg} \\
Impact threshold $\varepsilon_i$ & $E_g$ to $W$~\cite{OkutoCrowell1974} \\
\bottomrule
\end{tabular}
\end{table}

\subsection{Chynoweth calibration details (methods)}
\label{subsec:chynoweth_methods}

\paragraph*{Goal and observable.}
For a calibration diode that produces an approximately uniform electric field over a known effective multiplication width \(d\), the (initiating-carrier) ionization coefficient can be inferred directly from the measured gain,
\begin{equation}
\alpha_{\mathrm{init}}(E,T)\;=\;\frac{1}{d}\ln M(V), 
\end{equation}
\begin{equation}
M(V)\equiv \frac{\text{measured signal gain at bias }V}{\text{baseline (no-ICA) gain}}.
\end{equation}
In the onset regime, \(\alpha_{\mathrm{init}}(E,T)\) is well described by a Chynoweth form,
\(\alpha_{\mathrm{init}}(E,T)=A(T)\exp[-B(T)/E]\), so a plot of \(\ln \alpha_{\mathrm{init}}\) versus \(1/E\) is approximately linear:
\begin{equation}
\ln \alpha_{\mathrm{init}} \;=\; \ln A \;-\; \frac{B}{E},
\Rightarrow\quad \text{slope}=-B,\;\; \text{intercept}=\ln A.
\end{equation}
A key cryogenic cross-check is that the fitted slope magnitude decreases upon cooling, i.e.\ \(B(4~\mathrm{K})<B(77~\mathrm{K})\), consistent with longer inelastic free flights at lower temperature. If the opposite trend is observed, it typically indicates temperature instability, an inaccurate field mapping \(E(V)\), or an unaccounted device-heating/leakage artifact.

\paragraph*{Calibration and injection polarity.}
To extract carrier-specific coefficients and maintain \emph{unipolar} conditions, we define the initiating carrier as the carrier drifting \emph{into} the highest-field region and choose the bias polarity so that this carrier experiences the dominant ionization over the effective width \(d_{\mathrm{eff}}\). In a planar diode with a well-defined multiplication layer, reversing the bias swaps the initiating carrier: one polarity yields \emph{hole-initiated} multiplication and thus constrains \(\alpha_h(E,T)\), while the opposite polarity yields \emph{electron-initiated} multiplication and constrains \(\beta_e(E,T)\). Where practical, this separation is strengthened using near-surface optical injection: short-wavelength illumination from the high-field-side electrode localizes photocarrier generation near the contact and makes the injection dominated by the carrier drifting into the multiplication layer, enabling independent Chynoweth fits for \(\alpha_h\) and \(\beta_e\). For point-contact/ICA-style devices, the same principle is implemented through contact choice and field shaping to suppress bipolar feedback (e.g., favoring the initiating carrier with the larger ionization coefficient in Ge), and the extracted \((A(T),B(T))\) parameters are validated by reproducing measured \(M(V)\) curves together with the corresponding Chynoweth plots.

\paragraph*{Device geometry and $d$ metrology.}
We recommend short, \emph{uniform-field} test diodes (e.g., planar p$^+$–i–n with a-Ge/Li contacts) whose
multiplication region is a well-controlled slab of width \(d\). The effective \(d\) used in \(\alpha=(\ln M)/d\)
must be measured at the \(\mu\)m scale:
(i) process-calibrated etch/implant depth,
(ii) stylus/optical profilometry of the recess,
(iii) cross-sectional SEM/TEM on witness samples, and
(iv) TCAD field maps to confirm uniformity. Quote \(d\) with an uncertainty (e.g., \(d=5.0\pm0.3~\mu\mathrm{m}\))
and propagate it into \(\alpha\).

\paragraph*{Temperature stability.}
Measurements are performed at fixed cryogenic \(T\) (77~K or 4~K) with active control
(stability better than \(\pm 20\)–\(50\) mK) to avoid drift in mobility and inelastic scattering.
Log temperature and hold time at each bias step.

\paragraph*{Front-end gain and noise.}
Calibrate the electronics gain with a precision test pulser (or known X-ray line) at low bias (\(M\simeq 1\)).
Record the equivalent noise charge (ENC) and bandwidth. For each \(V\), estimate \(M(V)\) from pulse-height ratios
corrected for any bandwidth changes. Exclude points where the SNR is \(<10\) or where pile-up/heating affects the mean pulse height.

\paragraph*{Field estimate $E(V)$.}
Convert bias to field using (in increasing fidelity):
(i) a 1D depletion model \(E\approx (V-V_{\mathrm{dep}})/d_{\mathrm{eff}}\) with a focusing factor \(\eta\) inferred from TCAD;
(ii) full 2D/3D TCAD (Poisson–drift–diffusion) using measured leakage and contact boundary conditions to extract the \emph{local} field in the slab; or
(iii) a hybrid approach that maps the measured depletion curve to a calibrated \(E(V)\).
Use the same $E(V)$ model across the dataset and report its uncertainty (e.g., \(\pm 5\%\)).

\paragraph*{Fit window and regression.}
Identify the near-onset region where \(\ln\alpha\) vs \(1/E\) is linear (typically a decade in \(\alpha\) above the detection limit but below pre-breakdown curvature).
Use robust linear regression (e.g., Huber) or ordinary least squares with outlier rejection; report the fitting window \([E_{\min},E_{\max}]\) and \(R^2\).
Extract \((A,B)\) and the covariance matrix \(\Sigma\); confidence intervals follow from standard errors or nonparametric bootstrap over \((M,V)\).

\paragraph*{Uncertainty propagation.}
Propagate uncertainties from \(d\), \(E(V)\), and the fit into \(\alpha\) and \((A,B)\). For device-level figures of merit,
\[
E_{\mathrm{crit}}^{(\mathrm{PI})} \;=\; \frac{B}{\ln(Ad)}\,,
\]
a convenient linearized propagation is
\[
\sigma^2\!\left(E_{\mathrm{crit}}^{(\mathrm{PI})}\right)
= \left(\frac{\partial E_{\mathrm{crit}}}{\partial A}\right)^2 \sigma_A^2\]
\[+ \left(\frac{\partial E_{\mathrm{crit}}}{\partial B}\right)^2 \sigma_B^2\]
\[+ 2\,\frac{\partial E_{\mathrm{crit}}}{\partial A}\frac{\partial E_{\mathrm{crit}}}{\partial B}\,\mathrm{Cov}(A,B),
\]
with \(\partial E_{\mathrm{crit}}/\partial A = -B/[A(\ln Ad)^2]\) and \(\partial E_{\mathrm{crit}}/\partial B = 1/\ln(Ad)\).
Confidence bands on \(M(V)\) follow from sampling \((A,B)\) consistent with the fitted covariance and re-evaluating \(M=\exp[\alpha(E(V))d]\).

\subsubsection*{Worked example (synthetic dataset)}
\begin{itemize}
  \item \textbf{Geometry:} planar test diode, uniform multiplication slab \(d=5.0\pm0.3~\mu\mathrm{m}\).
  \item \textbf{Conditions:} 77~K, temperature stability \(\pm 30\) mK; electronics SNR \(>20\) across window.
  \item \textbf{Field map:} TCAD-derived \(E(V)\) with an estimated \(\pm 5\%\) systematic.
  \item \textbf{Data window:} \(\alpha d\in[0.05,\,1.0]\) (i.e., \(M\in[1.05,\,2.72]\)); linear \(\ln\alpha\)–\(1/E\) behavior with \(R^2=0.994\).
\end{itemize}
Linear regression of \(\ln\alpha\) on \(1/E\) yields
\[
A_{77K}=(3.2\pm0.3)\times 10^{5}~\mathrm{cm^{-1}},\]
\[B_{77K}=(3.9\pm0.4)\times 10^{4}~\mathrm{V/cm}
\]
(1\(\sigma\) uncertainties; covariance \(\mathrm{Cov}[\ln A,\,B/E]\approx -0.12\)).
For \(d=5~\mu\mathrm{m}\), the resulting device-onset estimate is
\[
E_{\mathrm{crit}}^{(\mathrm{PI})}(77\ \mathrm{K},d)=\frac{B_{77}}{\ln(A_{77}d)}
=\frac{3.9\times10^{4}}{\ln(3.2\times10^{5}\times 5\times10^{-4})}\]
\[\approx\;7.6\times 10^{3}\ \mathrm{V/cm},
\]
with a propagated \(\pm 12\%\) (1\(\sigma\)) uncertainty dominated by \(B\) and \(\delta d\).
Repeating the procedure at 4~K (same device, new dataset) should yield a \emph{smaller}
slope parameter than at 77~K. An illustrative outcome is
\[
A_{4\mathrm{K}}=(4.1\pm0.4)\times10^{5}~\mathrm{cm^{-1}},\]
\[B_{4\mathrm{K}}=(2.9\pm0.3)\times10^{3}~\mathrm{V/cm},
\]
and
\(E_{\mathrm{crit}}^{(\mathrm{PI})}(4~\mathrm{K},5~\mu\mathrm{m})\approx 4.8\times10^{2}\ \mathrm{V/cm}\),
consistent with longer inelastic free flights at 4~K.

\paragraph*{Illustrative values (electrons, uniform field).}
To make the scaling concrete, we quote \emph{plausible} transport numbers consistent with cryogenic Ge trends and use them to compute
\(E_{\mathrm{crit}}^{(\mathrm{PI})}\) for two multiplication widths. For electrons, take at 77~K:
\(A_{77K}\!=\!3.0\times 10^{5}~\mathrm{cm^{-1}}\), \(B_{77K}\!=\!4.0\times 10^{4}~\mathrm{V/cm}\);
and at 4~K (updated to enforce the physically expected trend \(B(4~\mathrm{K})<B(77~\mathrm{K})\)):
\(A_{4K}\!=\!4.0\times 10^{5}~\mathrm{cm^{-1}}\), \(B_{4K}\!=\!2.8\times 10^{3}~\mathrm{V/cm}\).
Using \cref{eq:EcritPI} gives:
\begin{table}[h]
\centering
\caption{Illustrative PI onset fields for electrons in Ge (uniform field).}
\label{tab:PI_examples}
\begin{tabular}{lccc}
\toprule
\(T\) & \(d\) & \(\ln\!\big(A\,d\big)\) & \(E_{\mathrm{crit}}^{(\mathrm{PI})}\) (V/cm) \\
\midrule
77 K & \ (10\,$\mu$m) & \(5.704\) & \(\approx 7.0\times 10^{3}\) \\
77 K & \ (5\,$\mu$m)  & \(5.011\) & \(\approx 8.0\times 10^{3}\) \\
4 K  & \ (10\,$\mu$m) & \(5.991\) & \(\approx 4.8\times 10^{2}\) \\
4 K  & \ (5\,$\mu$m)  & \(5.298\) & \(\approx 5.3\times 10^{2}\) \\
\bottomrule
\end{tabular}
\end{table}

\noindent
Two points are worth emphasizing:
(i) the \emph{logarithmic} dependence on \(d\) means that halving \(d\) raises \(E_{\mathrm{crit}}\) only modestly;
(ii) the temperature dependence enters primarily through \(B(T)\), so reduced inelastic scattering at 4~K can shift the onset field substantially~\cite{Jacoboni1983}.
For comparison, the SFF upper bound at 77~K with \(m_e^\ast\!=\!0.12\,m_0\), \(\varepsilon_i\!\approx\!2.96\) eV, and \(\mu_e(77~\mathrm{K})\!\approx\!3.6\times 10^4~\mathrm{cm^2/V\,s}\)
is \(E_{\mathrm{crit}}^{(\mathrm{SFF})}\!\approx\!8.1\times 10^{3}\) V/cm (Sec.~\ref{sec:sff}). The PI values in \cref{tab:PI_examples} are therefore
\emph{comparable to or slightly below} the SFF bound, as expected once \(\ln(A d)\) and the calibrated inelastic physics are included. At 4~K, the reduced \(B(T)\)
moves the PI onset to lower field, consistent with the enhanced multiplication expected in cryogenic Ge.
Note that, because direct mobility measurements at 4~K are not available, the estimate above relies on the \emph{effective} mobilities inferred from SuperCDMS drift-velocity data at 30--50~mK. If, in practice, the effective mobility at 4~K is \emph{not} an order of magnitude larger than the mobility at 77~K---for example, because neutral-impurity and defect scattering continue to limit transport---then the critical field \(E_{\mathrm{crit}}^{(\mathrm{PI})}\) is expected to remain close to the 77~K value, consistent with the behavior discussed by Wei \textit{et al.}~\cite{Wei2022PPPCGe}.

\paragraph*{Using the calibrated model for design.}
With \((A,B)\) in hand, one can:
\begin{itemize}
  \item \textbf{Map operating windows.} Convert \(E_{\mathrm{crit}}^{(\mathrm{PI})}(T,d)\) to onset voltages \(V_{\mathrm{B}}=E_{\mathrm{crit}}d\) for candidate geometries and compare to leakage limits from I--V data or TCAD.
  \item \textbf{Trade gain vs.\ noise.} Combine \(\alpha(E)\) with McIntyre excess noise \(F(k,M)\) (Sec.~\ref{sec:physics-informed}) to identify bias/gain points that meet a target energy resolution (minimize \(k=\beta/\alpha\) via field shaping and contact engineering)~\cite{McIntyre1966,OkutoCrowell1974}.
  \item \textbf{Cross-check against SFF.} Use \cref{eq:SFF} as a quick upper bound for new corners; large discrepancies between SFF and PI predictions often indicate nonuniform fields or missing inelastic channels.
\end{itemize}

\paragraph*{Visualization.}
The SFF mobility/threshold scaling is summarized in \cref{fig:Ecrit_vs_mu}, the multiplication width \(d\) and field localization are shown in \cref{fig:ica_schematic} and \cref{fig:field}, and the experimental extraction of \((A,B)\) is organized naturally through the Chynoweth representation in \cref{fig:chynoweth}.

\section{Discussion and outlook}
\label{sec:discussion}

The preceding sections culminate in a calibrated, device-facing workflow:
(i) use mobility and a SFF bound to set fast, conservative targets
(Sec.~\ref{sec:sff}); (ii) replace the one-flight picture with a
physics-informed (PI) ionization law \(\alpha(E,T)=A(T)e^{-B(T)/E}\) that
encodes nonparabolic kinematics and energy-dependent inelasticity
(Sec.~\ref{sec:physics-informed}); (iii) translate \(\alpha(E,T)\) into an
onset field through the device-level criterion
\(\int_0^d\alpha(E,T)\,dx\simeq 1\) (Sec.~\ref{sec:device_level}); and
(iv) tie these ingredients to a specific Ge ICA geometry and TCAD field map
(Sec.~\ref{sec:device_design}) and to a practical calibration recipe with
uncertainty bands (Sec.~\ref{sec:calibration}). Here we place this
framework in context, summarize the dominant design levers for cryogenic Ge
ICA detectors, and outline near-term experimental and modeling priorities.

Many of the underlying building blocks used here—lucky-drift arguments, the
Chynoweth/Okuto--Crowell form for the impact-ionization coefficient, and
McIntyre-type treatments of avalanche noise—are standard in the
avalanche-photodiode (APD) literature and have been developed in great
detail for Si and Ge devices over several decades.~\cite{Conwell1967,Jacoboni1983,McIntyre1966,Okuto1975}
Full-band Monte Carlo simulations and semi-empirical Chynoweth fits are
routinely used to describe high-field transport and breakdown in
conventional APDs, while low-temperature detector communities often rely
on empirical onset fields inferred from device-specific conditioning curves.

The present work differs from these approaches in three ways that are
directly tailored to cryogenic Ge detectors and the Ge ICA architecture.
First, we make explicit the bridge from the SFF picture to a PI
energy--relaxation integral, linking \cref{eq:SFF} to
Eqs.~(\ref{eq:energyrelaxing})--(\ref{eq:AB}) and thereby connecting the
Chynoweth slope parameter \(B(T)\) to microscopic velocity and inelastic
scattering. This provides a transparent mapping between textbook
lucky-drift arguments and the temperature-dependent \(\alpha(E,T)\) used
in design. Second, we show that for a finite multiplication region of
width \(d\) the PI framework yields a compact design rule
\(
E_{\mathrm{crit}}^{(\mathrm{PI})} = B(T)/\ln\!\big(A(T)d\big)
\)
(\cref{eq:EcritPI}), which can be viewed as a transport-informed
refinement of the SFF reference scale and is portable across geometries and
operating temperatures. Third, we demonstrate how the calibrated
\(\alpha(E,T)\) can be embedded into realistic Ge ICA modeling by
combining it with TCAD electric-field profiles and measured impurity
gradients, enabling breakdown margins, multiplication factors, and the
impact of field nonuniformity to be assessed for a specific detector
geometry (\cref{fig:ica_schematic,fig:field}). In this sense, our goal is
not to replace full-band Monte Carlo tools, but to provide a lightweight,
physics-based modeling layer that captures the dominant cryogenic
avalanche physics in a form directly usable by Ge detector designers.

\paragraph*{Dominant design levers.}
The unified framework yields compact formulas with clear physical handles:
mobility \(\mu(T)\) sets the SFF scale (\cref{eq:SFF}); nonparabolicity and
energy-dependent inelastic spectra set the Chynoweth slope \(B(T)\)
(\cref{eq:AB}); post-threshold physics sets the prefactor \(A(T)\)
(\cref{eq:alpha}); and geometry enters only logarithmically through
\(\ln\!\big(A(T)d\big)\) in \cref{eq:EcritPI}. Importantly, the PI model
predicts a \emph{smaller} slope parameter at lower temperature:
suppressed phonon absorption and reduced inelastic relaxation at
\(4~\mathrm{K}\) increase the effective inelastic free flight, which lowers the
energy-relaxation integral defining \(B(T)\) and therefore shifts
\(\alpha(E,T)\) upward (larger \(\alpha\) at fixed \(E\)). Hence
\(B(4~\mathrm{K})<B(77~\mathrm{K})\) and the onset field
\(E_{\mathrm{crit}}^{(\mathrm{PI})}\) decreases at 4~K for a fixed \(d\),
all else equal.~\cite{Jacoboni1983,OkutoCrowell1974}
Because \(\ln(A d)\) varies slowly, the most powerful levers for lowering onset
are (i) reducing \(B(T)\) through temperature and purity (suppressing phonon-
and defect-assisted inelasticity), and (ii) enforcing unipolar-favored
multiplication to minimize excess noise \(F(k,M)\) with \(k\) controlled
via field shaping and contact technology.~\cite{SzeNg,McIntyre1966,OkutoCrowell1974,Jacoboni1983}
The device schematic in \cref{fig:ica_schematic} emphasizes how field shaping
(anode size, guard gap, passivation, bevel) localizes the high-field region of
effective width \(d\), while \cref{fig:Ecrit_vs_mu} summarizes conservative SFF
trends against achievable cryogenic mobilities.

\paragraph*{Predicted multiplication versus bias for the Fig.~\ref{fig:ica_schematic} design.}
For a first design pass, we connect the calibrated transport to a simple device map.
Let the multiplication region beneath the point contact be \(d=5~\mu\mathrm{m}\)
(\(5\times10^{-4}\)~cm) and adopt illustrative 77~K transport parameters
consistent with the PI trends: \(A_{77}=3\times10^{5}~\mathrm{cm^{-1}}\),
\(B_{77}=4\times10^{4}~\mathrm{V/cm}\).~\cite{OkutoCrowell1974,Jacoboni1983}
At 4~K, the temperature-consistent PI expectation is a modestly changed
prefactor but a \emph{smaller} slope parameter; for example,
\(A_{4\mathrm{K}}=4\times10^{5}~\mathrm{cm^{-1}}\) and
\(B_{4\mathrm{K}}=2.8\times10^{3}~\mathrm{V/cm}\), which implements
\(B(4~\mathrm{K})<B(77~\mathrm{K})\) and therefore yields larger \(\alpha\) (and
larger gain) at the same local field. We take a linear field-focusing map
between the measured depletion voltage and operating bias of the concept device
(\cref{fig:ica_schematic}): \noindent\textbf{(1) 77~K.}
We take $V_{\mathrm{dep}}\approx 3.2~\mathrm{kV}$ and $V_{\mathrm{op}}\approx 5.1~\mathrm{kV}$, and choose the proportionality constant in a linear bias-to-field map such that the local field at $V_{\mathrm{op}}$ matches the physics-informed onset value for $d=5~\mu\mathrm{m}$, i.e.,
\[
E(V_{\mathrm{op}})\simeq E_{\mathrm{crit}}^{(\mathrm{PI})}\approx 8\times10^{3}~\mathrm{V/cm}
\quad \text{(cf.\ Table~\ref{tab:PI_examples}).}
\]
The corresponding multiplication is
\[
M(V)=\exp\!\bigl[\alpha(E(V),T)\,d\bigr],\qquad
\alpha(E,T)=A(T)\exp\!\left(-\frac{B(T)}{E}\right),
\]
with
\[
E(V)=E(V_{\mathrm{op}})\,\frac{V-V_{\mathrm{dep}}}{V_{\mathrm{op}}-V_{\mathrm{dep}}}\,,\qquad V\ge V_{\mathrm{dep}},
\]
and $M\simeq 1$ for $V<V_{\mathrm{dep}}$.

\noindent\textbf{(2) 4~K.}
At 4~K we assume that no depletion voltage is required because impurity dopants are largely frozen out into localized neutral states, leading to negligible free-carrier space charge in the bulk~\cite{Mei2024ResidualImpuritiesGeDM}. To quantify how little the residual space charge perturbs the near-contact field, we evaluate the electric-field variation within a characteristic distance \(d=5~\mu\mathrm{m}\) of the point contact (PC). Since the PC radius (\(r_{\mathrm{pc}}=1~\mathrm{mm}\)) is orders of magnitude larger than this region, the local boundary is effectively planar on the \(5~\mu\mathrm{m}\) scale (curvature corrections are \(\mathcal{O}(d/r_{\mathrm{pc}})\)). For a representative residual ionized-impurity concentration \(N_a=2\times10^{8}~\mathrm{cm^{-3}}\), Gauss's law gives a space-charge-induced field change across \(d\),
\[
\Delta E \simeq \frac{q N_a}{\epsilon_{\mathrm{Ge}}}\,d,
\]
which evaluates to \(\Delta E \approx 1\times10^{-2}~\mathrm{V/cm}\) for \(d=5~\mu\mathrm{m}\), far smaller than the bias-driven focusing field near the readout electrode. Thus, the near-contact field is set primarily by the applied bias and the weighting-potential geometry rather than by the bulk impurity profile. In this high-gradient PPC/ICPC-like regime, it is convenient to approximate the localized high-field region beneath the point contact as an effective uniform slab of thickness \(d\) with a representative field \(E_{\mathrm{eff}}\). For order-of-magnitude estimates, \(E_{\mathrm{eff}}\) may be related to the applied bias through a hemispherical/near-spherical scaling \(E\sim \eta V_{\mathrm{bias}}/r_{\mathrm{pc}}\), where \(\eta=\mathcal{O}(0.1\text{--}1)\) accounts for departures from an ideal spherical field in a semi-infinite detector bulk~\cite{Cooper2011NovelHPGe,Abt2007GERDASegmented,Mertens2019PPC}. In general, TCAD-based \(E(\mathbf{r})\) maps are preferred to capture the full nonuniform field distribution and to identify potential failure modes such as geometric hot spots and microplasma-prone regions. However, for 4~K operation, a fully predictive TCAD treatment is not yet available because the requisite cryogenic charge-transport microphysics (e.g., temperature-dependent scattering and impact-ionization parameterizations) in Ge is not sufficiently established for robust device-level simulation; we therefore adopt simplified field-mapping assumptions for the 4~K illustrative calculations.

For the illustrative 4~K $M(V)$ curves below, we therefore adopt a simple linear field-focusing map referenced to a starting bias \(V_{\mathrm{start}}=150~\mathrm{V}\) and a reference operating bias \(V_{\mathrm{ref}}=400~\mathrm{V}\), with the target local field
\[
E(V_{\mathrm{ref}})=5.3\times10^{2}~\mathrm{V/cm}\quad \text{(Table~IV).}
\]
Accordingly,
\[
E(V)=E(V_{\mathrm{ref}})\,\frac{V-V_{\mathrm{start}}}{V_{\mathrm{ref}}-V_{\mathrm{start}}}\,,\qquad V\ge V_{\mathrm{start}},
\]
and \(M(V)\) is computed from the same expression as above.

\begin{figure}[t]
  \centering
  \includegraphics[width=0.92\linewidth]{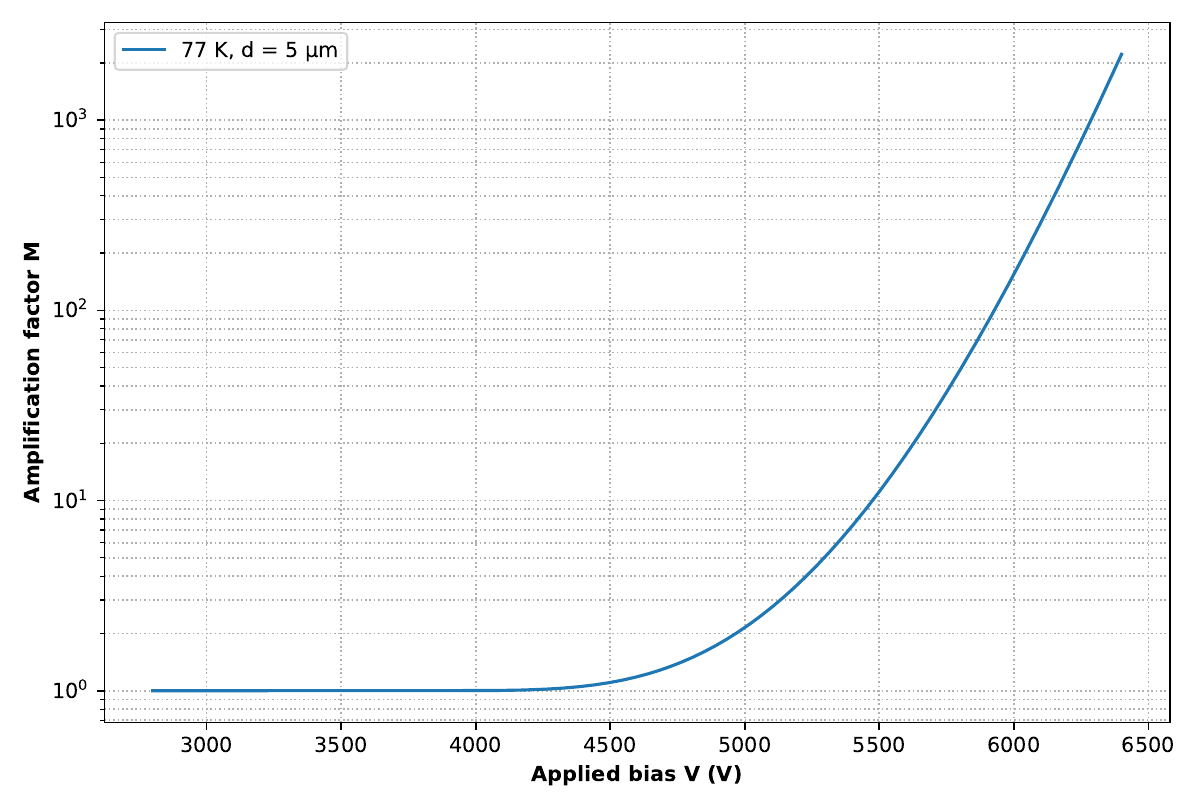}
  \includegraphics[width=0.92\linewidth]{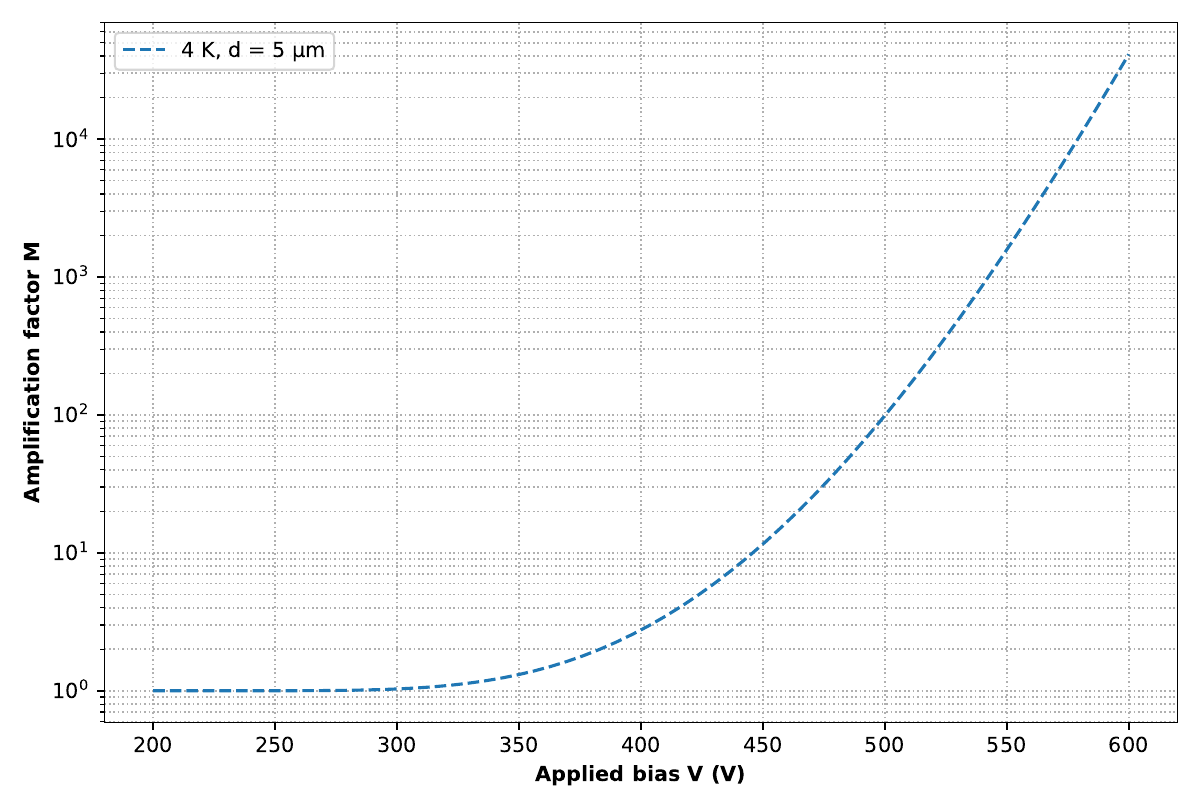}
 \caption{\textbf{Predicted amplification $M(V)$ for the ICA concept (Fig.~\ref{fig:ica_schematic}) at 77~K and 4~K.}
The top panel shows the predicted amplification at 77~K and the bottom panel shows the predicted amplification at 4~K.
We use an illustrative impact-ionization parameterization $\alpha(E,T)=A(T)\exp[-B(T)/E]$ with
$A_{77}=3\times10^{5}~\mathrm{cm^{-1}}$, $B_{77}=4\times10^{4}~\mathrm{V/cm}$ and
$A_{4\mathrm{K}}=4\times10^{5}~\mathrm{cm^{-1}}$, $B_{4\mathrm{K}}=2.8\times10^{3}~\mathrm{V/cm}$,
and a multiplication width $d=5~\mu\mathrm{m}$.
The device field is mapped linearly from bias.
At 77~K, we assume $E(V_{\mathrm{dep}})=0$ at $V_{\mathrm{dep}}=3.2~\mathrm{kV}$ and
$E(V_{\mathrm{op}})=8\times10^{3}~\mathrm{V/cm}$ at $V_{\mathrm{op}}=5.1~\mathrm{kV}$.
At 4~K, no depletion voltage is assumed; instead we map the field linearly from
$V_{\mathrm{start}}=150~\mathrm{V}$ with $E(V_{\mathrm{start}})=0$ to
$E(V_{\mathrm{ref}})=5.3\times10^{2}~\mathrm{V/cm}$ at $V_{\mathrm{ref}}=400~\mathrm{V}$.
The smaller $B$ at 4~K (longer inelastic mean free paths) yields larger $M$ at the same bias.
In practice, TCAD and measured diode $M(V)$ data will be used to refine $E(V)$ and calibrate $\{A(T),B(T)\}$.}

  \label{fig:MV_curve}
\end{figure}

The predicted \(M\)–\(V\)
characteristic (log–scale) is shown in \cref{fig:MV_curve}: gain rises gently
just above depletion and increases rapidly as \(V\) approaches the onset regime,
as expected from an exponential \(\alpha(E,T)\) law. Because \(B\) is smaller at
4~K, the 4~K curve lies above the 77~K curve at the same bias, reflecting longer
inelastic free flights and enhanced ionization probability. In practice,
nonuniform fields and surface leakage (Sec.~\ref{sec:physics-informed}) soften
this curve and may require derating of \(V_{\mathrm{op}}\), which is best
quantified with TCAD and \(M(V)\) measurements on uniform test diodes.~\cite{SzeNg,Jacoboni1983}

\begin{figure}[t]
  \centering
  \includegraphics[width=0.92\linewidth]{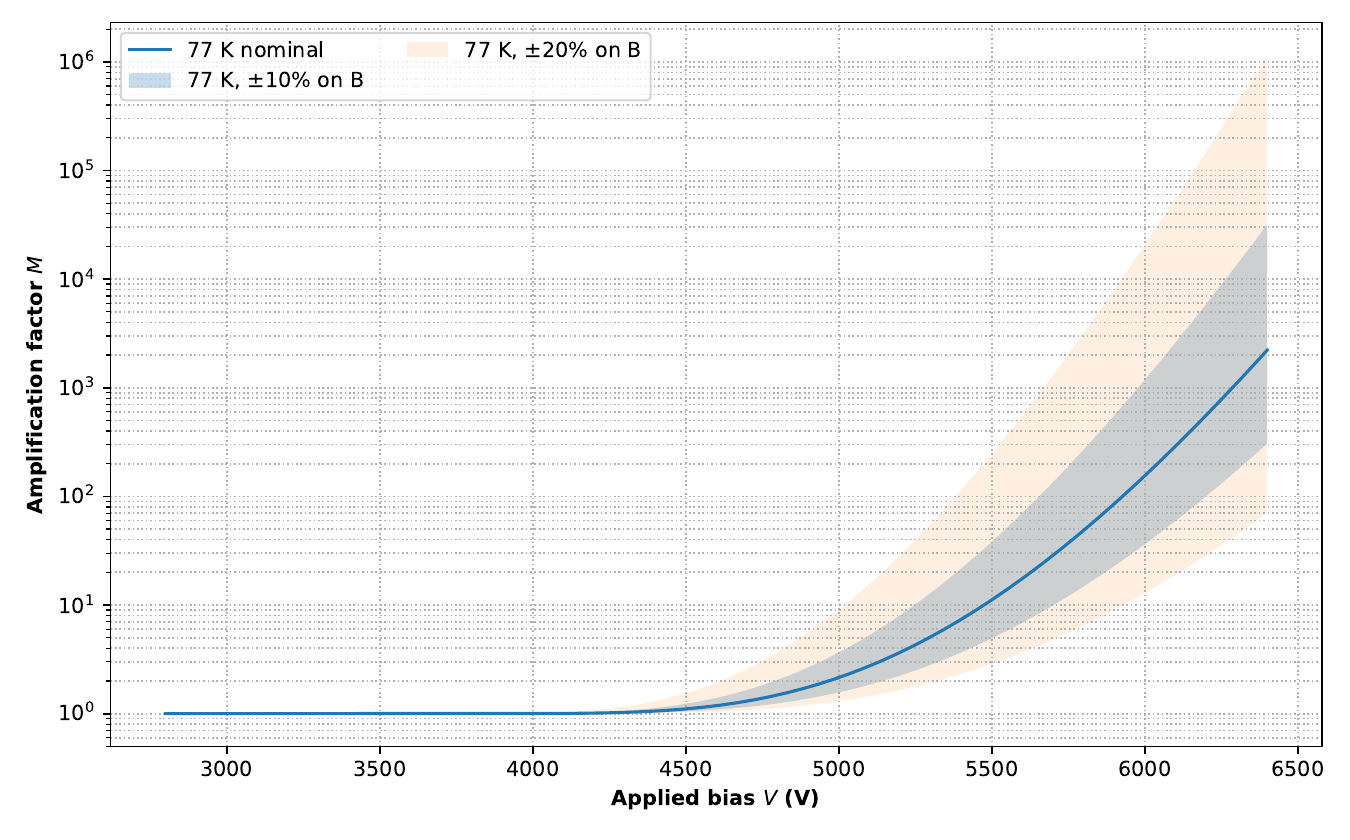}
  \includegraphics[width=0.92\linewidth]{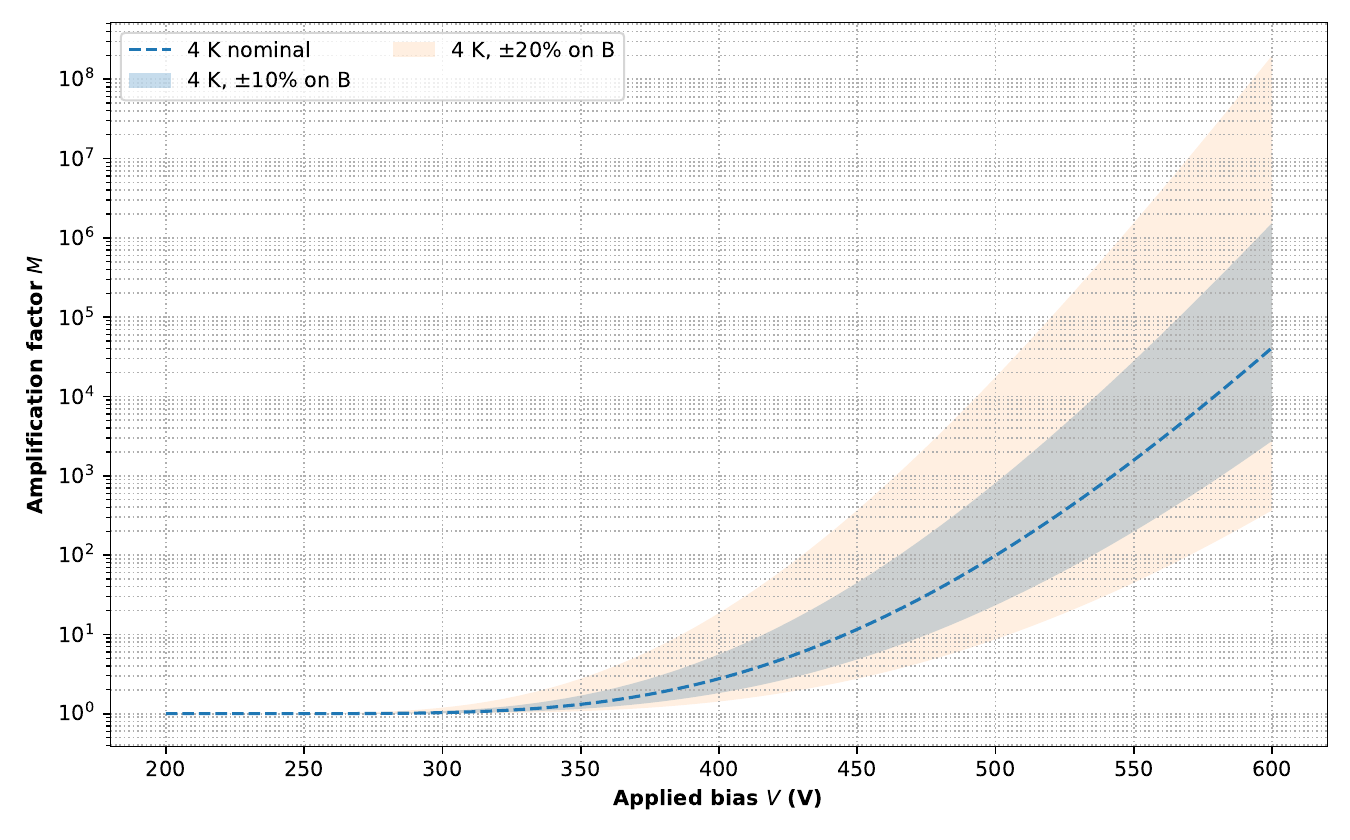}
  \caption{\textbf{Robustness of $M(V)$ to transport uncertainty.}
Predicted amplification $M(V)$ for the ICA concept assuming a uniform multiplication region of thickness
$d=5~\mu\mathrm{m}$, shown for 77~K (top) and 4~K (bottom).
The impact-ionization coefficient is modeled as $\alpha(E,T)=A(T)\exp[-B(T)/E]$ using nominal $(A,B)$ values from
Sec.~\ref{sec:physics-informed}.
Shaded bands quantify the effect of uncertainty in $B$: $\pm10\%$ (darker) and $\pm20\%$ (lighter).
Below the depletion voltage $V_{\mathrm{dep}}$, we set $M=1$.
The smaller $B$ at 4~K increases $\alpha$ at fixed $E$, shifting $M(V)$ upward at the same bias.
These uncertainty bands complement the SFF reference scale and the physics-informed estimate
$E_{\mathrm{crit}}^{(\mathrm{PI})}=B/\ln(Ad)$ (\cref{eq:EcritPI}), and should be refined using measured diode
$M(V)$ data and TCAD-derived $E(V)$ maps~\cite{OkutoCrowell1974,Jacoboni1983,SzeNg}.}
\label{fig:MV_uncertainty}
\end{figure}

To assess robustness, Fig.~\ref{fig:MV_uncertainty} shows the effect of
\(\pm 10\%\) and \(\pm 20\%\) variations in \(B\) at 77~K and 4~K.
Because \(B\) controls the exponential sensitivity of \(\alpha(E)\), the spread
in \(M(V)\) is largest in the high-field portion of the curve. Even so, the
banded prediction demonstrates that a calibrated \((A(T),B(T))\) pair can set
first-pass bias targets; improved TCAD-derived \(E(V)\) maps and reduced
uncertainty on \(d\) will further tighten these bands. We emphasize that the
curves in Figs.~\ref{fig:MV_curve} and \ref{fig:MV_uncertainty} are illustrative
and use plausible \((A,B)\) values rather than parameters obtained from a direct
fit to data.

\paragraph*{Noise/readout linkage (quantitative example).}
The multiplication curves above are only useful if gain does not
sacrifice resolution via excess avalanche noise. Using McIntyre’s model
for the excess-noise factor~\cite{McIntyre1966},
\(F(k,M)=k\,M+(2-1/M)(1-k)\), one can translate a target gain into an
effective improvement in the input-referred noise. For a unipolar-favored
geometry (\(k\ll 1\)), \(F\simeq 2-1/M\) so that \(F(5)\approx 1.8\) and
\(F(10)\approx 1.9\). A practical rule-of-thumb for input-referred
equivalent noise charge is then
\[
\mathrm{ENC}_{\mathrm{eff}}\;\approx\;\frac{\mathrm{ENC}_{0}}{M}\,\sqrt{F},
\]
so a cryogenic front-end with \(\mathrm{ENC}_{0}=60~e^{-}_{\mathrm{rms}}\)
would improve to \(\mathrm{ENC}_{\mathrm{eff}}\approx 17~e^{-}\) at \(M=5\)
and \(\approx 8.3~e^{-}\) at \(M=10\). In energy units, using
\(\varepsilon_{\mathrm{pair}}\approx 2.96~\mathrm{eV}\) for Ge at cryogenic
\(T\),
\[
\mathrm{FWHM}_{E}\;\approx\;2.355\,\varepsilon_{\mathrm{pair}}\,\mathrm{ENC}_{\mathrm{eff}},
\]
giving \(\mathrm{FWHM}_{E}\approx 118~\mathrm{eV}\) at \(M=5\) and
\(\approx 58~\mathrm{eV}\) at \(M=10\), compared to \(\approx 418~\mathrm{eV}\)
at \(M=1\). If multiplication becomes more bipolar (e.g., \(k=0.1\)),
\(F(5)\approx 2.12\) and \(F(10)\approx 2.71\), degrading the examples to
\(\mathrm{FWHM}_{E}\approx 123~\mathrm{eV}\) and \(\approx 69~\mathrm{eV}\),
respectively—quantitatively reinforcing why Ge ICA designs should favor
unipolar-favored field shaping and contacts.

\paragraph*{Outlook and next steps.}
Two near-term validation loops are natural. First, fabricate and measure
short, uniform-field Ge test diodes with the same contact stack as the
target detector to extract \((A(T),B(T))\) at 77~K and 4~K from Chynoweth
plots (\(\ln\alpha\) vs \(1/E\)). A decisive cryogenic validation is that the
extracted slope satisfies \(B(4~\mathrm{K})<B(77~\mathrm{K})\) over the onset
window; if not, revisit temperature control, \(E(V)\) mapping, and the \(d\)
metrology. Second, run calibrated TCAD (Poisson–drift–diffusion with SRH traps
and surface boundary conditions) to convert \(V\) to the relevant local \(E(x)\),
identify hot-spots and microplasma-prone regions, and project \(M(V)\) and
\(F(k,M)\) under realistic fields.~\cite{ShockleyRead1952,Hall1952,SzeNg}
Longer term, incorporating valley anisotropy, refined phonon spectra, and
benchmarks against full-band Monte Carlo will tighten the mapping from
microscopic transport to device gain and noise. The same calibrated
\(\alpha(E,T)\) layer can then be ported to alternative Ge detector
geometries (ring-contact, strip, and segmented designs) where deliberate
field localization is used to confine multiplication while maintaining
low leakage and stable operation.

\section{Conclusion}
\label{sec:conclusion}

Building on the calibration and device-design examples in
Sec.~\ref{sec:discussion}, we have presented a compact, data-anchored
framework for predicting the critical electric field
\(E_{\mathrm{crit}}\) required to initiate ICA in high-purity Ge at 77~K and 4~K. The approach unifies a
mobility-based SFF bound with a PI
impact-ionization model that explicitly embeds
(i) energy-dependent inelastic scattering,
(ii) nonparabolic carrier dispersion (Kane bands),
(iii) intervalley transfer, and
(iv) the high-energy (``lucky drift'') tail of the carrier distribution.~\cite{SzeNg,Kane1957,Jacoboni1983,Conwell1967,Chynoweth1958,OkutoCrowell1974}
In the parabolic/constant-\(\tau\) limit the PI model reduces to the SFF
result \cref{eq:SFF}; with realistic transport, it yields the exponential
ionization law \(\alpha(E,T)\simeq A(T)\exp[-B(T)/E]\)
(\cref{eq:alpha,eq:AB}) and, through the device-level onset criterion, the
closed-form design rule
\(
E_{\mathrm{crit}}^{(\mathrm{PI})}=B(T)/\ln[A(T)\,d]
\)
(\cref{eq:EcritPI}), which connects microscopic transport to the effective
multiplication width \(d\).

\paragraph*{What the formulas enable.}
These relations provide a practical toolbox for detector design rather than
a purely formal re-derivation. First, \cref{eq:SFF} ties onset fields
directly to the measured low-field mobility \(\mu(T)\), making the
temperature and purity lever arm explicit through
\(E_{\mathrm{crit}}\propto \mu^{-1}\).~\cite{SzeNg}
Second, \cref{eq:alpha,eq:AB} package band structure and inelastic
spectra into the Chynoweth parameters \((A,B)\), which can be extracted
\emph{directly} from uniform-field diode measurements via
\(\ln\alpha\)–\(1/E\) (Chynoweth) plots.~\cite{Chynoweth1958,OkutoCrowell1974}
Third, \cref{eq:EcritPI} folds \((A,B)\) into geometry, highlighting the
modest (logarithmic) leverage of \(d\) and the dominant role of transport
and temperature in setting onset. In particular, the expected cryogenic
trend \(B(4~\mathrm{K})<B(77~\mathrm{K})\) (longer inelastic free flights)
implies a lower PI onset field at 4~K for a fixed \(d\), shifting
multiplication to lower bias compared to 77~K.~\cite{Jacoboni1983}
Figures~\ref{fig:Ecrit_vs_mu} and \ref{fig:ica_schematic}  translate these
dependencies into concrete design guidance by emphasizing field
localization under a small anode and providing quick targets versus
achievable mobilities at cryogenic temperature. The resulting illustrative
\(M(V)\) characteristics (Fig.~\ref{fig:MV_curve}) show the expected
gentle-onset/rapid-rise behavior near breakdown that follows from an
exponential \(\alpha(E)\).

\paragraph*{Implications for low-threshold detectors.}
For low-threshold instrumentation, the framework clarifies how ICA can
raise signal above front-end noise while adding only modest excess noise
when multiplication is effectively unipolar (\(k=\beta/\alpha\ll1\)).
In that regime, the same compact relations inform co-design of bias,
geometry, and contact technology to minimize McIntyre’s excess-noise
factor \(F(k,M)\) at the desired gain.~\cite{McIntyre1966,OkutoCrowell1974,SzeNg}
At 4~K, suppressed phonon absorption lengthens inelastic free flights,
reducing \(B(T)\) and hence \(E_{\mathrm{crit}}\) relative to 77~K—an
advantage for applications requiring sub-keV (and potentially sub-100~eV)
thresholds, including low-mass dark matter searches and CE\(\nu\)NS.~\cite{Jacoboni1983,Mei2018EPJC}
The explicit temperature dependence of \(A(T)\) and \(B(T)\) provides a
quantitative handle for setting bias margins and acceptable heat loads in
cryogenic systems.

\paragraph*{Limits and validation pathway.}
At the same time, the SFF and PI relations assume a locally uniform field
and do not by themselves capture space-charge and surface-current
feedback. In real devices, surface leakage, dead layers, SRH traps, and
field hot-spots (curvature or roughness) can lower the usable bias and
trigger microplasma before the mean field reaches
\(E_{\mathrm{crit}}^{(\mathrm{PI})}\).~\cite{SzeNg}
To close this gap, we advocate a two-step validation loop:
(i) extract \((A,B)\) from short, uniform-field test diodes at 77~K and
4~K via \(M(V)\) and Chynoweth plots, verifying the expected decrease in
the Chynoweth slope \(B(T)\) at 4~K; and
(ii) deploy calibrated TCAD (Poisson–drift–diffusion with SRH
recombination and realistic surface boundary conditions) to convert bias
\(V\) into \(E(x)\), locate hot-spots, and predict multiplication and noise
under realistic fields.~\cite{ShockleyRead1952,Hall1952,SzeNg,Jacoboni1983}
Agreement between PI-predicted \(E_{\mathrm{crit}}\) and observed onsets in
well-behaved diodes would validate the transport parameters and de-risk
scale-up to ICA geometries such as point-contact, ring-contact, or strip
detectors.

\paragraph*{Outlook and experimental path.}
The framework is lightweight enough for early design optimization yet
sufficiently grounded in transport physics to capture the strong cryogenic
trends expected in Ge. Immediate theory extensions include incorporating
explicit L-valley anisotropy, benchmarking the PI-derived \(\alpha(E,T)\)
against full-band Monte Carlo, and porting the methodology to other
semiconductors relevant to cryogenic rare-event detection. For HPGe ICA,
the closed-form relations
\cref{eq:SFF,eq:alpha,eq:AB,eq:EcritPI} together with
Figs.~\ref{fig:ica_schematic}, \ref{fig:Ecrit_vs_mu}, and \ref{fig:MV_curve}
provide actionable guidance to set target mobilities, bias margins, and
multiplication widths that meet gain/noise goals for LDM and CE\(\nu\)NS
instrumentation.

From a practical standpoint, this framework is already being used to guide
the first generation of Ge ICA prototypes. The initial devices under
fabrication target a gain window of \(M \sim 20\text{--}50\) at 77~K, with a
high-field region of effective width \(d \simeq 5~\mu\mathrm{m}\) and
operating biases in the \(4\text{--}6~\mathrm{kV}\) range; these choices
follow directly from the PI-based onset criterion
\(E_{\mathrm{crit}}^{(\mathrm{PI})}=B/\ln(Ad)\) and the SFF upper bound
developed in Sec.~\ref{sec:physics-informed}.
In the near term, we will validate the model using short test diodes and
simplified Ge ICA structures, measuring multiplication curves \(M(V)\),
breakdown-field distributions across devices, and excess-noise factors
\(F(M)\) as a function of temperature. Such data will enable refined
extraction of \(\alpha(E,T)\) and \((A(T),B(T))\), test the expected
cryogenic reduction of \(B(T)\) at 4~K, close the loop between
the PI transport layer and device-level observables, and provide the
empirical grounding needed for subsequent, higher-gain Ge ICA detectors and
related cryogenic Ge sensors.

\begin{acknowledgments}
We acknowledge support from the National Science Foundation under Grants
No.~OISE-1743790, PHYS-2117774, OIA-2427805, PHYS-2310027, and OIA-2437416,
and from the U.S.\ Department of Energy under Grants
No.~DE-SC0024519 and DE-SC0004768, as well as support from a research center
funded by the State of South Dakota. We also thank our Ge-STAR collaborators Joel Sander, Juergen Reichenbacher, and Xinhua Bai
for helpful discussions and for their careful review of this manuscript.
\end{acknowledgments}

\bibliographystyle{apsrev4-2}
\bibliography{ge-avalanche}

\end{document}